%% file: composition.tex
\documentclass{aa}
\pdfoutput=1
\usepackage{graphicx}
\usepackage[varg]{txfonts}
\usepackage{natbib}
\bibpunct{(}{)}{;}{a}{}{,} 
\usepackage{float}
\usepackage[usenames,dvipsnames]{xcolor}
\usepackage[figuresright]{rotating}
\newcommand{\new}[1]{#1}

\begin{document}
\title{The white dwarf's carbon fraction as a secondary parameter of Type Ia supernovae}
\titlerunning{Carbon fraction as a secondary parameter of SNe Ia}

\author{Sebastian~T.~Ohlmann\inst{1}\thanks{\email{sohlmann@astro.uni-wuerzburg.de}} \and
       Markus~Kromer\inst{2,3} \and
       Michael~Fink\inst{1} \and
       R\"{u}diger~Pakmor\inst{4} \and \\
       Ivo~R.~Seitenzahl\inst{1,3,6,7} \and
       Stuart~A.~Sim\inst{5,6,7} \and
       Friedrich~K.~R\"{o}pke\inst{1}
       }

\authorrunning{S.~T.~Ohlmann et al.}

\institute{%
    Institut f\"{u}r Theoretische Physik und Astrophysik,
    Universit\"{a}t W\"{u}rzburg, Emil-Fischer-Str. 31, 
    97074 W\"{u}rzburg, Germany
  \and
    The Oskar Klein Centre \& Department of Astronomy,
    Stockholm University, AlbaNova, SE-106 91 Stockholm, Sweden
  \and
    Max-Planck-Institut f\"{u}r Astrophysik, Karl-Schwarzschild-Str. 1, 
    85741 Garching, Germany
  \and
    Heidelberger Institut f\"{u}r Theoretische Studien,
    Schloss-Wolfsbrunnenweg 35, 69118 Heidelberg, Germany
  \and
    Astrophysics Research Centre, School of Mathematics and Physics, 
    Queen's University Belfast, Belfast BT7 1NN, UK
  \and
    Research School of Astronomy and Astrophysics, Mount Stromlo Observatory,
    Cotter Road, Weston Creek, ACT 2611, Australia
  \and
    ARC Centre of Excellence for All-sky Astrophysics (CAASTRO)
}


\abstract
  {
  Binary stellar evolution calculations predict that Chandrasekhar-mass
  carbon/oxygen white dwarfs (WDs) show a radially varying profile for the
  composition with a carbon depleted core. Many recent multi-dimensional
  simulations of Type Ia supernovae (SNe~Ia), however, assume the
  progenitor WD has a homogeneous chemical composition.
  } {
  In this work, we explore the impact of different initial carbon profiles of
  the progenitor WD on the explosion phase and on synthetic observables in the
  Chandrasekhar-mass delayed detonation model. Spectra and light curves are
  compared to observations to judge the validity of the model.
  } {
  The explosion phase is simulated using the finite volume supernova code
  \textsc{Leafs}, which is extended to treat different compositions of
  the progenitor WD\@. The synthetic observables are computed with the Monte
  Carlo radiative transfer code \textsc{Artis}.
  } {
  Differences in binding energies of carbon and oxygen lead to a lower nuclear
  energy release for carbon depleted material; thus, the burning fronts that
  develop are weaker and the total nuclear energy release is smaller.  For
  otherwise identical conditions, carbon depleted models produce less $^{56}$Ni.
  Comparing different models with similar $^{56}$Ni yields shows lower kinetic
  energies in the ejecta for carbon depleted models, but only small differences
  in velocity distributions and line velocities in spectra.  The light curve
  width--luminosity relation (WLR) obtained for models with differing carbon
  depletion is roughly perpendicular to the observed WLR, hence the carbon mass
  fraction is probably only a secondary parameter in the family of SNe~Ia.
  } {} 

\keywords{supernovae: general -- hydrodynamics -- nucleosynthesis -- radiative transfer
  -- binaries: close -- white dwarfs -- methods: numerical}

  \maketitle

\section{Introduction}

  Despite enormous efforts in recent years, both in observation and modeling,
  identification of the progenitors of Type Ia supernovae (SNe~Ia) remains
  elusive; no system has been observed so far
  \citep[e.g.,][]{roelofs2008a,li2011b} and model predictions do not allow us to
  unambiguously distinguish different progenitor scenarios
  \citep[e.g.,][]{roepke2012a}. It is generally agreed that SNe~Ia result from
  thermonuclear explosions of carbon/oxygen white dwarfs (WDs) in interacting
  binary systems, where mass transfer from the secondary component triggers the
  explosion \citep[e.g.,][]{hillebrandt2000a}. Depending on whether the secondary
  star is degenerate, the systems are distinguished in single degenerate (SD)
  systems --~with main sequence, red giant, or sdB companion stars, for example~--
  and double degenerate (DD) systems --~with He or carbon/oxygen (C/O) WDs as
  secondary components.  For a review of explosion models see
  \citet{hillebrandt2000a}; more recent explosion models and their comparison
  to observations are presented in \citet{hillebrandt2013a}.  Interestingly,
  for the subclass of SN~2002cx-like SNe, pure deflagrations of
  Chandrasekhar-mass \citep{chandrasekhar1931a} WDs
  \citep{jordan2012b,kromer2013a,fink2014a} in the SD channel match observables
  quite well \citep{kromer2013a}.  For normal SNe~Ia, the most promising
  candidates are delayed detonations of Chandrasekhar-mass WDs in SD systems
  \citep[e.g.,][]{golombek2005a,gamezo2005a,roepke2007b,bravo2008a,jordan2008a,
    townsley2009a,bravo2009a,bravo2009b,jackson2010a,seitenzahl2011a,roepke2012a,
  jordan2012a,seitenzahl2013a,sim2013a}, double detonations of
  sub-Chandrasekhar-mass WDs in SD and DD systems
  \citep[e.g.,][]{fink2007a,fink2010a,moll2013a}, and violent mergers of
  sub-Chandrasekhar-mass WDs in DD systems
  \citep[e.g.,][]{pakmor2012b,pakmor2012a,pakmor2013a,moll2013b,raskin2013a}.  A
  recent comparison to SN~2011fe \citep{roepke2012a} shows that disentangling
  different models is hard given the current predictions in the optical regime.
  Other suggestions for progenitor scenarios include head-on collisions
  \citep[e.g.,][]{rosswog2009a,raskin2009a,kushnir2013a,garcia2013a} and the
  core-degenerate scenario \citep[e.g.,][]{soker2014a}.

  The level of interest in SNe~Ia has risen in the past 20 years mainly because
  they can be used as distance indicators in cosmology (for a review, see
  \citealp{leibundgut2008a}), leading to the discovery of an accelerated
  expansion of the Universe
  \citep[e.g.,][]{riess1998a,schmidt1998a,perlmutter1999a}.  The foundation for
  this development was the establishment of a light curve width--luminosity
  relation (WLR, also called the Phillips relation) by \citet[see also the
  earlier work of \citealp{pskovskii1977a}]{phillips1993a}. This relation
  enables the estimation of absolute luminosities from the light curve width:
  broader light curves correspond to brighter SNe. 
  \new{Thus, to a first approximation, SNe~Ia are a one-parameter family driven
  by the \emph{primary} parameter which affects light curve width and
  luminosity in such a way that the observed WLR emerges. As observed SNe~Ia
  show some scatter around this WLR, some \emph{secondary} parameters have to
  be present influencing light curve width and luminosity following a
  relation different from the mean observed WLR. One challenge present in SN~Ia
  models is identifying the primary and secondary physical parameters (or sets
  of parameters) to better understand the physical origin of this relation.
  This should, as an ultimate goal, lead to a theoretical examination and
  justification of using this relation for a wide range of parameters, such as
  redshift or host stellar population.}

  In the multi-dimensional simulations presented here, we investigate the
  delayed detonation model in Chandrasekhar-mass WDs
  \citep{blinnikov1986a,khokhlov1991a}.  If Chandrasekhar mass models are to
  account for normal SNe~Ia \new{(in the sense of
  \citealp{branch1993a})}, the combination of a deflagration and a detonation
  is needed: neither pure detonation nor pure deflagration is
  sufficient. \new{Pure detonations
    of Chandrasekhar-mass WDs produce almost no IME because
    of the high densities ($\gtrsim 10^7$~g~cm$^{-3}$ in most of the WD;
    \citealp{arnett1971a}, see also introduction in \citealp{khokhlov1991a}).
    Pure deflagrations have recently been identified as promising models for
    2002cx-like SNe~Ia \citep{jordan2012b,kromer2013a}, but in other parameter
    ranges they fail to reproduce SNe~Ia
    \citep{fink2014a,ma2013a,roepke2008c}.}
  The densities through which the detonation propagates have to be lowered to
  produce more intermediate mass elements (IME). Hence, a deflagration flame
  first burns from the core outwards to the surface, thereby expanding the WD
  before the detonation is initiated.  The mechanism of igniting the detonation
  is unclear, but several possibilities are proposed and explored in 3D
  simulations: the spontaneous deflagration-to-detonation transition
  (spontaneous DDT,
  \citealp{gamezo2005a,roepke2007b,bravo2008a,townsley2009a,jackson2010a,
  seitenzahl2011a,roepke2012a,seitenzahl2013a,sim2013a}), the gravitationally
  confined detonation (GCD, \citealp{jordan2008a,jordan2012a}), and the
  pulsational reverse detonation (PRD, \citealp{bravo2009a,bravo2009b}).  These
  models differ in the way the detonation is initiated. \new{The detonation
    emerges only in the spontaneous DDT from regions where the flame turbulently
    mixes fuel with ashes; in the other models, the deflagration initiates a
    large-scale motion of gas, leading at some point to highly compressed hot spots,
    where the detonation is initiated. Thus, these models also differ in the
  hydrodynamical structure at the onset of the detonation.} 

  It is often assumed that for these models, the primary parameter of the WLR
  may be the ignition configuration, i.e., the shape of the initial deflagration
  flame (e.g., \citealp{seitenzahl2013a}). This parameter indeed leads to a
  variety of $^{56}$Ni masses and hence luminosities, but fails to reproduce the
  WLR in the recent study of \citet[although \citealp{kasen2009a} find
  their models populate a similar region to that of the WLR by changing ignition
  configuration and DDT criterion]{sim2013a}. 
  \new{Thus, recent multi-dimensional spontaneous DDT models
  \citep{sim2013a,seitenzahl2013a,roepke2012a} show reasonable agreement to
  observations, but some shortcomings remain, such as colors that are too red
  at maximum, velocities that are too high, and a failure to explain the observed
  WLR in terms of a sequence of models with differing initial deflagration
  strengths.}
  Therefore, it is vital to examine the consequences of other parameters.
  \new{For the single degenerate scenario considered here, the initial carbon
  fraction is one parameter that is expected to show variations for different
  progenitor systems depending on the zero-age main sequence (ZAMS) mass and on
  the metallicity of the progenitor star \citep{umeda1999b,dominguez2001a}.}
  It affects important parameters of the light
  curve evolution, such as the kinetic energy of the ejecta.  In 1D model studies,
  this has already been examined by \citet{hoeflich1998a} and \citet{umeda1999a};
  \citet{hoeflich2010a} also suggest the C fraction to be a secondary parameter.
  
  \new{The main objective of this work is to examine if varying the initial C
  fraction resolves any of the discrepancies between predictions of the
  spontaneous DDT model and observations of normal SNe Ia. More specifically, we
  want to answer the following two questions:}
  \begin{enumerate}
    \item \new{How does the carbon mass fraction affect the width--luminosity
      relation predicted for our 3D models? In particular, does it drive
      variations along the observed Phillips relation of SNe Ia?}
    \item Does a reduction of the carbon fraction result in better agreement of
      spectral features?
  \end{enumerate}

  To this end, we examine the impact of different initial
  compositions on the explosion process in multi-dimensional DDT models
  including its interplay with other parameters and present detailed
  nucleosynthesis results and synthetic light curves and spectra.  First, we
  give an overview of initial parameters involved in spontaneous DDT models in
  Section~\ref{sec:initialparameters}. In Section~\ref{sec:numericalmethods}, we
  explain our numerical methods: the initial WD models, the hydrodynamic
  modeling of the explosion phase, the detailed nucleosynthesis calculations,
  and the radiative transfer simulations. In Section~\ref{sec:resultshydro}, we
  examine the hydrodynamic evolution in a parameter study of 2D
  simulations as well as for a few 3D simulations and present
  detailed nucleosynthesis results. Results from radiative transfer simulations,
  synthetic light curves and spectra, for a series of 3D
  simulations are presented in Section~\ref{sec:resultsrt} along with a
  discussion of the width--luminosity relation. We conclude in
  Section~\ref{sec:conclusions} with answers to the questions posed above.

\section{Initial Parameters of Delayed Detonation Models for SNe~Ia}
\label{sec:initialparameters}

  The hydrodynamic evolution of the explosion phase in spontaneous DDT models
  is governed by several parameters. In some cases, these are poorly constrained and
  in others, they are constrained to vary in a certain range.
  
  In this work, we systematically explore the \emph{initial carbon fraction} of the
  pre-explosion WD, which depends on the evolution of the system.  In the phase prior to
  explosion, simmering sets in, changing the composition in the interior. This
  has been modeled \citep[e.g.,][]{lesaffre2006a}, but it is still difficult to
  include important effects such as the URCA process
  \citep{lesaffre2005a,lesaffre2006a}.  Nevertheless, these calculations show
  that the progenitor WD is composed of an outer layer of accreted material
  with an equal-by-mass composition of C and O, and an inner convective core
  with a lower C mass fraction. According to the calculations of
  \citet{lesaffre2006a}, this mass fraction and the size of the convective core
  is correlated to other parameters, such as the central density or
  metallicity. 
  \new{The central C mass fraction depends on the ZAMS mass and the metallicity
    of the progenitor \citep{umeda1999b,dominguez2001a}. The range in the
    central C fraction is about 0.24 to 0.37 for the models of
    \citet{umeda1999b} and about 0.21 to 0.32 for the models of
    \citet{dominguez2001a}. However, because of limitations in the modeling and in
    the nuclear data, the central C concentration is rather uncertain and
    \citet{dominguez2001a} suggest that it may vary between 0.1 and 0.5. The
    chemical stratification also changes during the simmering phase prior to
    ignition; therefore, we choose our model parameters similar to the models
    of \citet{lesaffre2006a}. We account for the possible spread in C
    fractions by varying it between 0.2 and 0.5 in our models.
  }
  The effect of the initial composition on the explosion is mainly
  by a different nuclear energy release in the burning: owing to the lower
  binding energy of C compared to O, material with less C possesses a smaller
  energy difference to the burning products, which are mostly in the iron group.

  Changing the \emph{initial metallicity} of the WD progenitor mainly
  influences the nucleosynthesis, since the principal influcence on the
  hydrodynamic evolution is due to the dependence of the equation of state on
  the electron fraction. In the nucleosynthesis, a lower metallicity and
  thus higher electron fraction leads to a larger $^{56}$Ni production
  \citep{timmes2003a,travaglio2005a,seitenzahl2013a} owing to reduced
  neutronization, but the effect is also not large enough for a primary
  parameter.

  The initial evolution of the deflagration flame is governed by the
  \emph{ignition configuration}, which is poorly understood.  Hydrodynamical
  simulations \citep{hoeflich2002a,kuhlen2006a,zingale2009a} favor a dipole
  structure of the convective flow preceding ignition, which may be fractured
  by rotation, yielding ignition over a broader region.  In our simulations, we
  place a certain number of ignition kernels near the center of the WD to
  excite different numerical modes of the deflagration flame. The number of
  these kernels determines the strength of the deflagration, i.e., the rate of
  energy production \citep{seitenzahl2013a,fink2014a}. More ignition kernels
  corresponds to a stronger deflagration and thus to a stronger expansion of
  the WD\@. Consequently, lower densities result at the onset of the detonation
  and thus, less $^{56}$Ni is produced. The range in deflagration strengths
  studied in \citet{seitenzahl2013a} is able to reproduce the observed range in
  brightnesses of normal SNe~Ia, but fails to explain the WLR \citep{sim2013a}.
  This study implies that the deflagration strength is probably not the primary
  parameter.

  Another important factor in the hydrodynamic evolution is the
  \emph{DDT criterion}. Although it is still unknown if and how this
  transition is realized in SNe Ia, some restrictions on the mechanism have been
  placed. The proposed instant of DDT to occur is when the flame leaves the
  flamelet regime and enters the distributed burning regime
  \citep{woosley2009a}, where hot ashes and cold fuel mix in the presence of
  large turbulent velocity fluctuations; hot spots result and a detonation is
  initiated.  \citet{woosley2007a} estimate the density of the DDT to lie in the
  range of (0.5--1.5)$\times10^{7}$~g~cm$^{-3}$.  The higher the density at the
  DDT is, the more $^{56}$Ni is produced (see the 2D simulations by
  \citealp{kasen2009a}). In the 3D simulations by \citet{seitenzahl2013a}, the
  DDT criterion was fixed for all simulations in order to examine the
  consequences of different ignition conditions. It is unclear why the DDT
  criterion should change in different SNe, if it is the primary parameter.

  The \emph{central density} of the WD at ignition is on the order of
  $10^9\,$g$\,$cm$^{-3}$.  However, some variation is possible, as the mass of a
  near-Chandrasekhar-mass WD is almost independent of the central density; this
  depends on the pre-explosion evolution. Higher initial densities shift the
  explosion products to more neutron-rich nuclei, mostly stable iron group
  elements (IGE) instead of $^{56}$Ni.  In their parameter study with 2D
  simulations, \citet{krueger2010a} find that a higher central density leads to
  similar overall production of IGE; but a lower amount of $^{56}$Ni is
  produced, as more neutron-rich nuclei are formed. \citet{seitenzahl2011a}, in
  contrast, find in their 3D simulations  a higher IGE production for higher
  central densities, whereas the amount of $^{56}$Ni is roughly constant. In any
  case, this effect is too small to be a primary parameter.

\section{Numerical Methods}
\label{sec:numericalmethods}

  To compute synthetic observables from explosion models, we use a modeling
  pipeline: after creating the initial WD models, the explosion phase is
  simulated using the hydrodynamic code \textsc{Leafs}; then, detailed
  nucleosynthesis results are computed in a postprocessing step; finally,
  synthetic observables are obtained with the radiative transfer Monte Carlo
  code \textsc{Artis}, using mapped data from the previous steps.

\subsection{Initial WD Models}

  The initial WD models have been created as cold isothermal WDs by integrating
  the hydrostatic equilibrium equations for a central density of
  $2.9\times10^9$~g~cm$^{-3}$ and a constant temperature of $5\times10^5$~K. The
  equation of state used for the integration is the same as in \textsc{Leafs}.
  The composition of the WD is chosen based on the results of
  \citet{lesaffre2006a}: uniform composition in the convective core and in the
  outer accretion layer with a smoothly connecting transition region. In the
  outer accretion layer, $X(^{12}\mathrm{C})=0.5$, whereas in the convective
  core $X(^{12}\mathrm{C})$ ranges from 0.2 to 0.5, depending on the model. The
  convective core ends at \new{about $1 M_\odot$} and the accretion layer starts at $1.2
  M_\odot$; again these values correspond to a typical scenario from
  \citet{lesaffre2006a}. \new{The size of the convective core depends on the
  chosen ignition criterion and for a certain choice of parameters, its mass is
  about $1 M_\odot$ for a central density of $2.9\times10^9$~g~cm$^{-3}$
  \citep[compare solid lines in fig. 7 of][]{lesaffre2006a}.}
  \new{We fix the mass of the convective core for all models and vary only the C
  fraction in order to have only one parameter changing in our models.
  Moreover, we take a rather wide range of the C fraction from 0.2 to 0.5 to
  assess the maximum possible influence of this parameter on the explosion
  process and on observables. Other findings indicate that the mass of the
  convective core may vary in a rather wide range, depending on variables such
  as the chemical stratification \citep{piro2008b} and uncertainties in the
  nuclear reaction rate data \citep{bravo2011b}. Moreover, in the models of
  \citet{bravo2011b}, a C depleted core develops in the innermost $0.05 M_\odot$
  of the WD. This should not influence the evolution of the flame much because
  the evolution of the flame in our 3D models is mostly governed by the
  Rayleigh--Taylor instabilities at later times. Thus, a C depletion in the very
  core should not significantly change the hydrodynamical evolution and thus
  also not the resulting observables.}

  To clarify the composition of each model, we introduce a naming scheme. The
  first part of the model names encodes the initial composition: cXY denotes a
  homogeneous progenitor model with a carbon mass fraction of XY\%.  Additional
  more realistic progenitor models \citep[following][]{lesaffre2006a} are
  labeled rpcXY corresponding to a homogeneous carbon depleted core with a
  carbon mass fraction of XY\%.

\subsection{Hydrodynamic Simulations}
\label{sec:leafs}
  We use the supernova code \textsc{Leafs} \citep{reinecke1999a,reinecke2002b}
  for the hydrodynamic simulations of the explosion phase. It employs a
  finite-volume, grid-based scheme in a Eulerian formulation of the piecewise
  parabolic method by \citet{colella1984a}, in the \textsc{Prometheus}
  implementation by \citet{fryxell1989a}. The Riemann solver is implemented
  according to \citet{colella1985a}, being capable of using a general convex
  equation of state. The equation of state is based on the Timmes equation of
  state \citep{timmes2000a}.

  Thermonuclear flames are modeled with the levelset method
  \citep{osher1988a} as described by \citet{reinecke1999a,reinecke2002b}. This
  approach approximates the flame front as a discontinuity, which burns the
  nuclear fuel instantaneously. The large difference in scales of several orders
  of magnitude between the flame width ($\sim$mm--cm) and the grid cell size
  ($\sim$km) justifies this approximation. Nuclear burning is treated in an
  approximate scheme, yielding the final composition directly behind the front.
  To track the energy release, a simplified composition is used including five
  pseudo-species, $\alpha$ particles, $^{12}$C, $^{16}$O, ``Mg'' (representing
  IME) and ``Ni'' (representing iron group elements, IGE); these approximate
  fuel and burning products from the different burning stages.  Nuclear
  statistical equilibrium (NSE) is treated approximately by adjusting the mass
  fractions of IGE and $\alpha$ particles to follow the energy release depending
  on density and temperature.  The detailed nucleosynthetic yields are computed
  in a postprocessing step using the method of tracer particles (see
  Section~\ref{sec:postprocessing}).  The tables giving the composition behind
  the level set depend on the density and composition of the unburnt fuel and
  have to be calculated once, prior to the simulations. This is done using an
  iterative calibration method similar to \citet{fink2010a}. The method is
  extended to different initial compositions to allow for varying compositions
  (for further details, see Appendix~\ref{sec:calibration}).  
  
  For the computational grid, we use the moving hybrid grid technique as
  described in \citet{roepke2006a}. An inner equidistant grid tracks the
  deflagration flame and expands into the outer, exponentially spaced grid as
  the deflagration evolves to allow for high resolution in the beginning. The
  deflagration burning takes place in the flamelet regime of turbulent
  combustion.  The effects of turbulence on unresolved scales are accounted for
  by a subgrid-scale model, which is used to compute turbulent velocity
  fluctuations below the grid scale. For 2D models, the subgrid-scale model by
  \citet{niemeyer1995b} is used, whereas for 3D models, a more elaborate model
  is employed as introduced by \citet{schmidt2006b,schmidt2006c}.

  The deflagration-to-detonation transition (DDT) is assumed to occur when the
  turbulent burning changes from the flamelet regime to the distributed burning
  regime \citep{woosley2009a}. Here, the internal flame structure is disturbed
  by turbulent eddies due to an increased flame width at lower densities. This
  leads to heat transfer from hot ashes to cold fuel
  \citep{niemeyer1997b,woosley2007a}, whereupon hot spots may form potentially
  initiating a detonation via the Zel'dovich gradient mechanism
  \citep{zeldovich1970a}. The flame widths necessary for this transition are
  reached in a density range of (0.5--1.5)$\times$10$^7\,$g$\,$cm$^{-3}$
  \citep{woosley2007a}. Furthermore, high turbulent velocity fluctuations of
  the order of $10^8\,$cm$\,$s$^{-1}$ must be present at the flame front
  \citep{lisewski2000b,woosley2009a}, which was found  in 3D deflagration models
  by \citet{roepke2007d}. For our 2D models, the DDT criterion is modeled as in
  \citet{kasen2009a}, but with differing parameters. A detonation is initiated
  in a cell if the density lies in a certain range and if the Karlovitz number
  Ka is larger than a given minimum value. Since $\mathrm{Ka} \propto
  (u')^{3/2}$ \citep[][Supp. Information]{kasen2009a}, where $u'$ denotes the
  turbulent velocity fluctuations below the grid scale, this criterion requires
  the turbulent velocity fluctuations to be above a certain threshold. In three
  dimensions, the DDT criterion is modeled as described in
  \citet{ciaraldi2013a}, but varying the parameters.   In this criterion, an
  effective flame surface is calculated by choosing cells in a certain density
  and fuel mass fraction range. This surface is additionally multiplied by the
  probability of large velocity fluctuations being present and it is required to
  exceed a critical value for at least half an eddy turnover time to ensure
  sufficient mixing between fuel and ashes.

  For the 2D simulations, a grid size of $512\times 1024$ cells in
  axial symmetry was chosen, corresponding to a spatial resolution of
  $1.06\,$km in the inner part at the beginning of the simulation. The 3D
  simulations are full star simulations and use a grid with $512^3$ cells,
  which corresponds to a spatial resolution of $2.14\,$km in the inner part at
  the beginning of the simulation.
  \begin{table}
    \centering
    \caption{Parameters for DDT criteria for 2D models similar to \citet{kasen2009a}. For details
    on the different parameters see ibid., supplementary information. The
    densities are given in $10^7\ \mathrm{g}\ \mathrm{cm}^{-3}$.}
    \begin{tabular}{c c c c}
      \hline\hline
      Criterion & $\rho_\mathrm{min}$ & $\rho_\mathrm{max}$ & Ka$_\mathrm{min}$ \\
      \hline
      ddt1 & 0.6 & 1.2 & 250 \\
      ddt2 & 0.5 & 0.8 & 1000 \\
      ddt3 & 0.5 & 0.8 & 2250 \\
      ddt4 & 0.6 & 1.2 & 2250 \\
      \hline
    \end{tabular}
    \label{tab:ddtcriteria2d}
  \end{table}

  The model names for 2D models consist of three parts; the first part gives the
  initial composition, as explained above. The second part of the model name
  consists of the DDT criterion; the corresponding parameters are given in
  Table~\ref{tab:ddtcriteria2d} and are similar to those used by
  \citet{kasen2009a}. The last part of the model name is determined by the
  initial conditions, the number gives the number of initial ignition spots for
  the deflagration flame.  The DDT criteria and ignition conditions are the same
  as in \citet{kasen2009a} with slightly different notations. The parameter
  study comprises of runs for five different initial composition profiles (c20,
  c30, c40, c50, and rpc32), for eight different ignition configurations (dd020,
  dd050, dd060, dd080, dd090, dd100, dd100C, and  dd150), and for two different
  DDT criteria (ddt1, ddt2). The rpc32 model has been run for all four DDT
  criteria of Table~\ref{tab:ddtcriteria2d}.

  For the 3D models, the treatment of initial composition is the same as for the
  2D models (see above). The values used for the limits in the DDT criterion are
  $0.4 < X_\mathrm{fuel} < 0.6$ and $0.6 < \rho / (\mathrm{g}\ \mathrm{cm}^{-3})
  < 0.9$ \citep[for details, see][]{ciaraldi2013a} \new{, where
    $X_\mathrm{fuel}$ is the mass fraction of unburnt material in the cell. The
    parameter range around 0.5 ensures that a detonation is ignited only in
    cells where fuel and ashes are mixed.}
  This criterion for the 3D models is termed DDT8 and differs from the one used
  by \citet{seitenzahl2013a}. The ignition conditions for the deflagration flame
  are the same as described by \citet{seitenzahl2013a}.
  
\subsection{Nucleosynthetic postprocessing}
\label{sec:postprocessing}

  Since coupling a reaction network to the hydrodynamic equations is
  computationally very expensive, we compute the detailed nucleosynthetic
  abundances in an additional postprocessing step. This was first done by
  \citet{thielemann1986a} for 1D models, computing a nuclear
  reaction network for the Lagrangian mass shells. For multi-dimensional
  simulations, we use the concept of \emph{tracer particles}, first introduced
  by \citet{nagataki1997a} in the context of Type II supernovae. In this method,
  tracer particles are passively advected with the flow and their thermodynamic
  trajectories are recorded. As the particles are moving in a Lagrangian manner,
  the nucleosynthetic abundances can be computed by evolving the nuclear network
  separately on each particle trajectory. The tracer particle method employed in
  our work is based on \citet{travaglio2004a} and uses the network of
  \citet{thielemann1996a} and \citet{iwamoto1999a} including 384 isotopes. More
  details on the algorithm can be found in \citet{roepke2006b}. 

  The distribution of the tracer particles is chosen according to
  \citet{seitenzahl2010a}, who proposed variable tracer masses in order to
  improve the resolution in the outer layers with lower densities.  The exact
  number of tracer particles depends on the simulation according to the
  algorithm by \citet{seitenzahl2010a} and is about $41000$ for 2D simulations
  and about $10^6$ for 3D simulations.

  The initial composition for the postprocessing is assumed to include the
  detailed solar metallicity of \citet{asplund2009a}. The CNO cycle elements
  are assumed to be processed to $^{22}$Ne during He burning; thus, their
  abundances are added by number to the $^{22}$Ne abundance.

\subsection{Radiative Transfer Simulations}

  The input data for the radiative transfer simulations is generated in the
  following way: the detailed nucleosynthesis data from the tracer particles is
  mapped onto a $200^3$ Cartesian grid using an SPH-like algorithm; the density
  distribution is mapped on this grid from the hydrodynamic simulation.  A
  further down-sampling to a $50^3$ grid yields the final input data for the
  radiative transfer calculation \citep[more details in][]{kromer2010a}.
  The radiative transfer simulations are then carried out with the multi-dimensional
  Monte Carlo code \textsc{Artis} \citep{sim2007b,kromer2009a}.
  On a co-expanding grid, following the homologous expansion of the ejecta,
  $10^8$ photon packages are propagated for 111
  logarithmically spaced time steps from 2~d to 120~d after explosion.
  The computations are sped up in the beginning by using a gray approximation in
  optically thick cells \citep[as discussed in][]{kromer2009a} and by assuming
  local thermodynamic equilibrium for the first 10 time steps, i.e., for the
  first two to three days post explosion. The atomic lines are taken from the
  same atomic data set as described in \citet{gall2012a}, including approximately
  $2\times 10^6$ bound--bound transitions. For the model N100 of
  \citet{seitenzahl2013a} and \citet{sim2013a}, the radiative transfer
  simulations have been recomputed with this large atomic data set.

\section{Hydrodynamic Evolution and Nucleosynthesis}
\label{sec:resultshydro}

  In this section, we present the results from the hydrodynamic simulations of
  the explosion phase and detailed nucleosynthetic abundances. This is first
  done for a set of 2D simulations, which can be run in larger numbers
  (compared to 3D simulations), owing to the lower computational effort. Then,
  the results for a few 3D simulations are presented.

\subsection{Parameter Study: 2D Simulations}

  A parameter study was performed in two dimensions to explore the impact of
  different initial compositions. To this end, hydrodynamical simulations of DDT
  models, followed by nucleosynthetic postprocessing, were performed for a set
  of five different initial compositions for a range of different ignition
  conditions and different DDT criteria. This also allows us to examine the
  effects of ignition conditions and DDT criteria separately and to compare
  these to the repercussions of the initial composition. Moreover, our parameter
  study results in models with similar $^{56}$Ni yields and thus similar
  luminosities; these may then be compared as models for the same supernova.

  \begin{figure}
    \centering
    \includegraphics{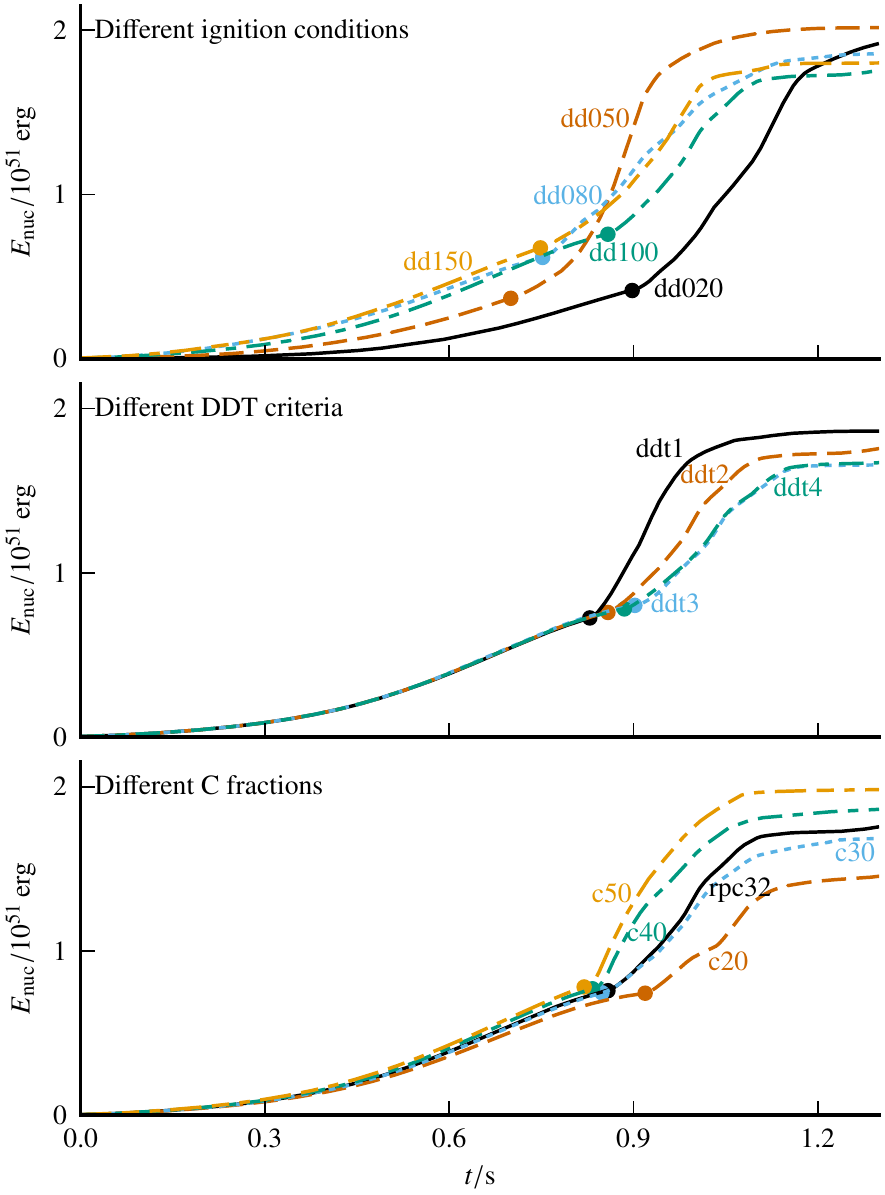}
    \caption{Evolution of nuclear energy release vs.\ time for several 2D
      simulations.  The dot marks the point of the first DDT\@. The upper panel
      shows simulations with different ignition conditions, but same initial
      composition and DDT criterion (rpc32\_ddt2\_ddxxx).  The middle panel
      shows simulations with different DDT criteria, but same initial
      composition and ignition conditions (rpc32\_ddtx\_dd03). The lower panel
      shows simulations with different C mass fraction, but same DDT criterion
      and ignition conditions (xxx\_ddt2\_dd03).
    }
    \label{fig:2devolution}
  \end{figure}
  \begin{figure}
    \centering
    \includegraphics{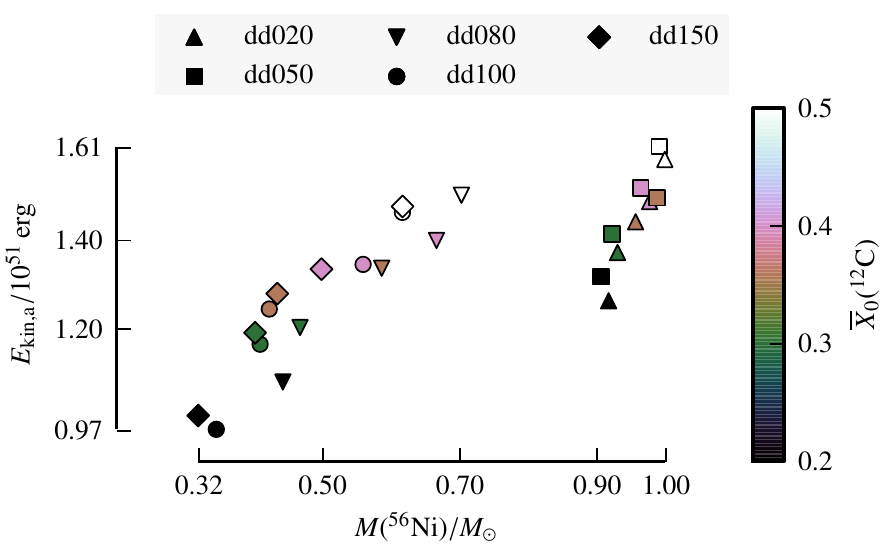}
    \caption{Final asymptotic kinetic energy over total $^{56}$Ni mass for
      several 2D models.  Different ignition configurations are shown with
      different markers. The average carbon mass fraction at the beginning of the
      simulation is color-coded. The DDT criterion is ddt2 for all models.
    }
    \label{fig:2dekinnimass}
  \end{figure}

  The repercussions of changing ignition conditions, DDT criteria, and initial
  compositions separately can be seen in Fig.~\ref{fig:2devolution}. In the
  upper panel, different \emph{ignition configurations} are compared. The
  nuclear energy release in the deflagration phase (prior to the first DDT
  marked with a dot) approximately increases with increasing number of ignition
  kernels. In this way, we can numerically excite varying deflagration strengths
  by changing the number of ignition kernels in our models. As the nuclear
  energy release is an indicator of the expansion of the WD, the WD has expanded
  more at the onset of the detonation for stronger deflagrations. Hence, for
  more ignition spots, the ensuing detonation burns less material at high
  densities; consequently, less $^{56}$Ni is produced and less nuclear energy is
  released in total \citep[see
  also][]{mazzali2007a,roepke2007b,seitenzahl2013a,fink2014a}. This relation,
  however, is fulfilled only approximately since the hydrodynamic evolution is
  highly non-linear and also depends on the locations of the ignition spots, not
  only on their number.

  Changing the \emph{DDT criterion} for otherwise identical conditions
  (Fig.~\ref{fig:2devolution}, middle panel) leads to a different delay until
  the detonation is ignited.  Later ignitions cause a lower $^{56}$Ni
  production and also a lower release of nuclear energy in total because of the
  longer-lasting pre-expansion.  
  
  The repercussions of the \emph{initial composition} on the hydrodynamic evolution
  for identical ignition configurations and DDT criteria can be seen in the
  lower panel of Fig.~\ref{fig:2devolution}. The homogeneous models show a
  slightly lower nuclear energy release in the deflagration phase for lower
  carbon mass fractions which can be explained by the lower energy release of
  deflagration fronts at lower carbon mass fractions (see
  Appendix~\ref{sec:calibration}).  This leads to a slower expansion and the
  turbulent velocity fluctuations needed for the DDT develop more slowly; thus,
  the detonation is initiated later.  This corresponds to a larger
  pre-expansion for lower carbon mass fractions. Hence less $^{56}$Ni is
  produced and less nuclear energy is released in total.  The more realistic
  model with a C depleted core (rpc32 model) is very similar to the homogeneous
  model with 30\% C in the deflagration phase because the deflagration flame
  does not leave the C depleted core. In the detonation phase, however, a
  larger nuclear energy release can be seen, which is due to the detonation
  burning also in the outer layers with larger C mass fractions.

  The nuclear energy release during the explosion phase drives the gravitational
  unbinding and the expansion of the ejecta; thus, the final, asymptotic kinetic
  energy of the ejecta is given by the sum of the initial internal energy, the
  initial gravitational energy (being negative) and the nuclear binding energy
  difference. This energy determines the scaling of the ejecta distribution in
  velocity space. The asymptotic kinetic energies and $^{56}$Ni yields are
  compared for several models in Fig.~\ref{fig:2dekinnimass}. First, models with
  larger $^{56}$Ni production show larger kinetic energies, which can be
  explained by the larger nuclear energy release. Second, when comparing
  models with identical ignition conditions but different initial compositions,
  a larger carbon mass fraction leads to a larger $^{56}$Ni yield and a larger
  asymptotic kinetic energy. The $^{56}$Ni masses for a model series with
  different initial compositions enclose a smaller interval for larger $^{56}$Ni
  masses because the detonation burns mostly at high densities, where the
  influence of the composition is small (see Appendix~\ref{sec:calibration}). 
  
\begin{figure}
    \centering
    \includegraphics{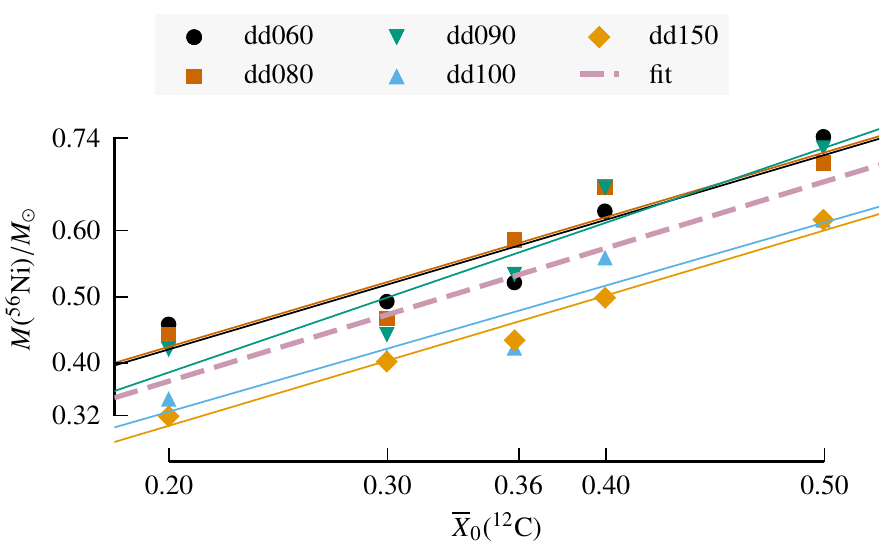}
    \caption{Total $^{56}$Ni mass over mean C mass fraction for several 2D
    models with different ignition conditions. The full lines show linear
    regressions for each ignition condition series; the dashed line shows the
    mean of all regressions.
    }
    \label{fig:2dnimasslinear}
  \end{figure}

  One important consequence is simply that different initial compositions lead
  to a spread in $^{56}$Ni masses for identical ignition conditions. The
  relation between the total $^{56}$Ni mass and the average initial C mass
  fraction $\overline{X}_0(^{12}\mathrm{C})$ is nearly linear
  (Fig.~\ref{fig:2dnimasslinear}). Linear regressions for all model series with
  $^{56}$Ni mass in the range for normal SNe~Ia ($\sim$0.3--0.8$M_\odot$) show
  correlation coefficients $> 0.94$. Averaging over these
  regressions yields an approximate expression for the $^{56}$Ni
  mass,
  \begin{equation}
    M(^{56}\mathrm{Ni})/M_\odot = 0.17 + 1.01 \overline{X}_0(^{12}\mathrm{C}),
    \label{eq:nimasslinear}
  \end{equation}
  for $0.2< \overline{X}_0(^{12}\mathrm{C}) <0.5$, showing a surprisingly simple
  mean relation between the initial C fraction and the $^{56}$Ni mass.
  
  If two different models are compared to the same SN, the models must
  produce a similar amount of $^{56}$Ni to show a similar luminosity. This can
  be reached by varying initial composition, ignition conditions, and DDT
  criterion at once. As can be seen in Fig.~\ref{fig:2dekinnimass}, the model
  with the smaller C mass fraction produces similar $^{56}$Ni yields for lower
  asymptotic kinetic energies than the model with the larger C mass fraction.
  Hence, the ejecta are distributed in a different way; and the light curves and
  spectra determined by this distribution will change.
  \begin{figure*}[p]
    \centering
    \includegraphics[width=0.95\textwidth]{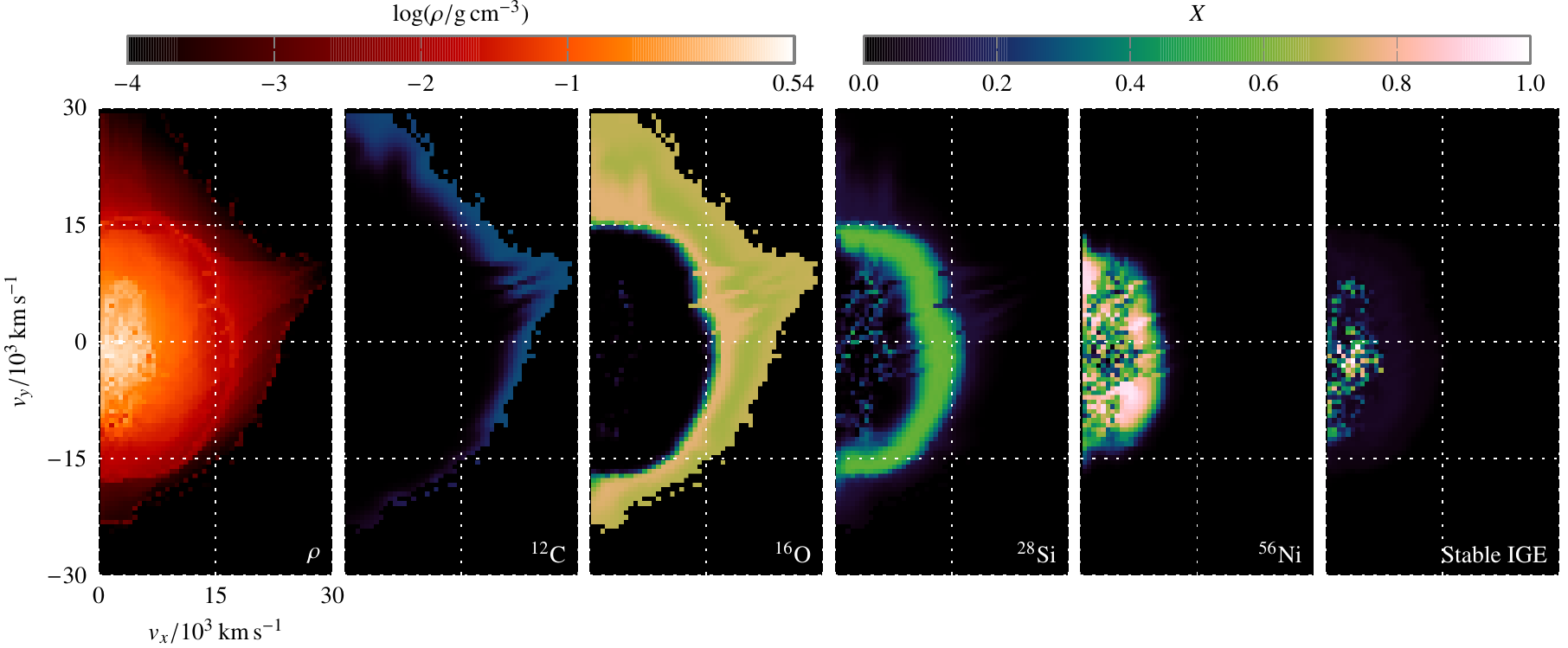}
    \caption{Ejecta distribution in velocity space for the 2D model
      c30\_ddt1\_dd150 producing $0.59\ M_\odot$ of $^{56}$Ni. Shown are the
      density (left panels) and mass fractions (right panels, stable IGE are all
      iron group elements with $Z>20$ without $^{56}$Ni).}
    \label{fig:velc30}
  \end{figure*}
  \begin{figure*}[p]
    \centering
    \includegraphics[width=0.95\textwidth]{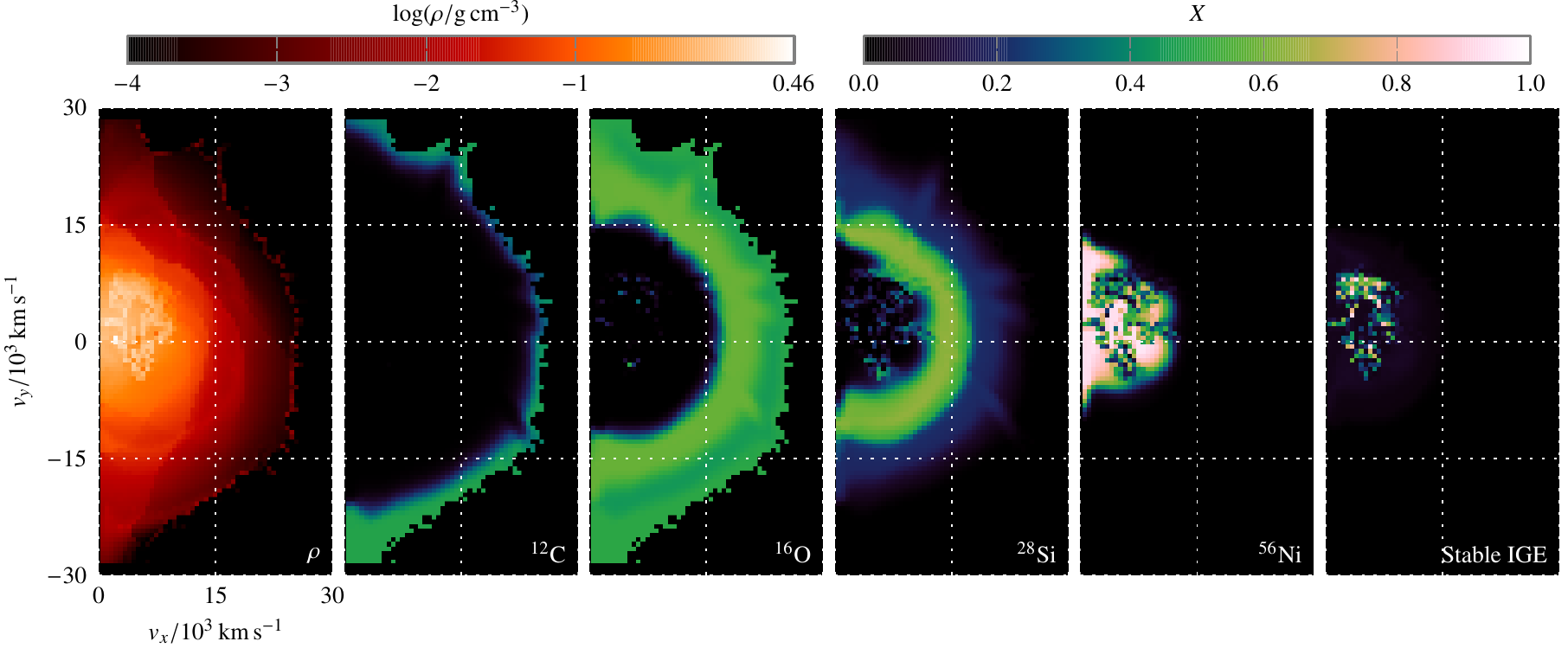}
    \caption{Same as Fig.~\ref{fig:velc30} for the 2D model
      rpc32\_ddt2\_dd080 producing $0.57\ M_\odot$ of $^{56}$Ni. 
      }
    \label{fig:velrpc32}
  \end{figure*}
  \begin{figure*}[p]
    \centering
    \includegraphics[width=0.95\textwidth]{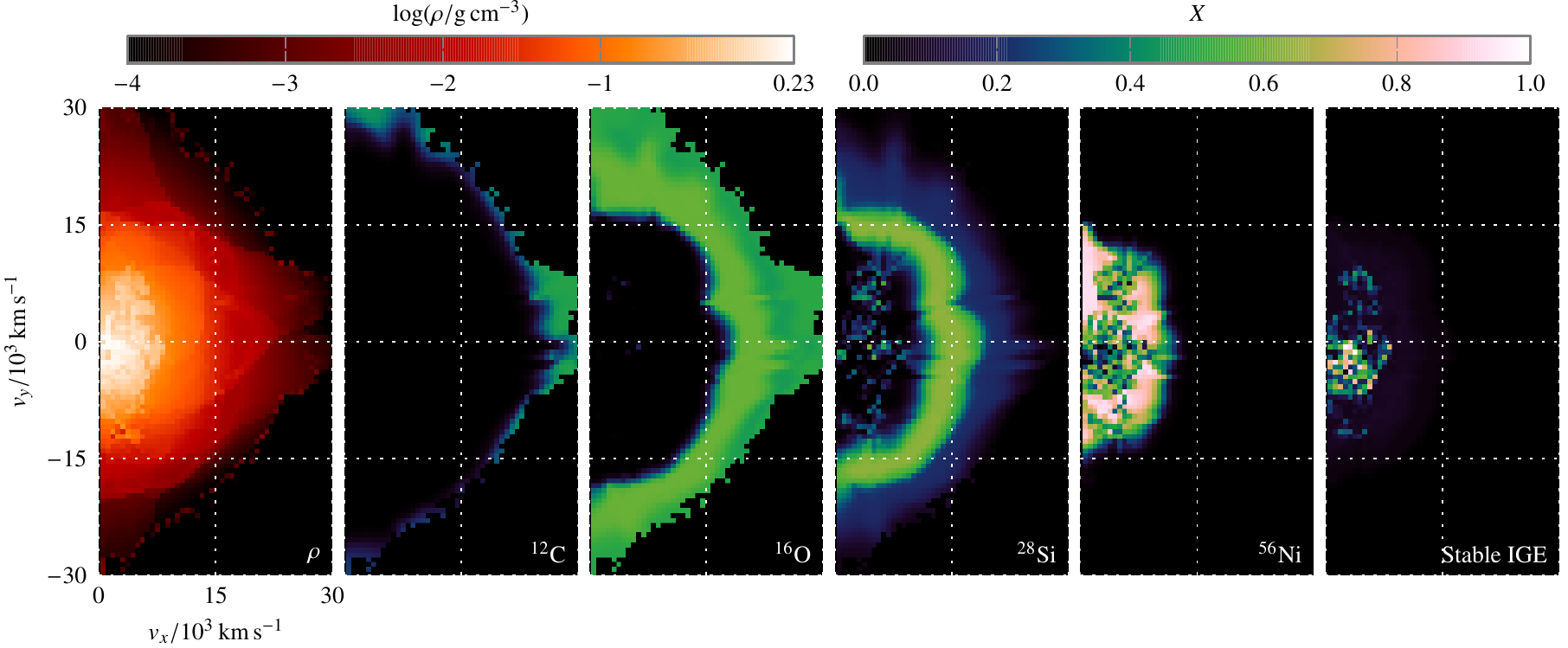}
    \caption{Same as Fig.~\ref{fig:velc30} for the 2D model
      c50\_ddt2\_dd150 producing $0.62\ M_\odot$ of $^{56}$Ni. 
      }
    \label{fig:velc50}
  \end{figure*}

  The ejecta distribution in velocity space is shown for three models with
  similar $^{56}$Ni masses but different initial compositions in
  Figs.~\ref{fig:velc30}, \ref{fig:velrpc32}, and \ref{fig:velc50}. 
  
  The density structure (left panel) shows similar features for all three
  models: a higher density in the interior part (${\la}10^4$~km~s$^{-1}$), where
  most deflagration ashes reside alongside detonation ashes; and shocks in the
  outer parts, where multiple detonation fronts merged.  The abundance structure
  (right panels) shows the same general features for all models: the central
  deflagration ashes are surrounded by layered detonation ashes. The models,
  however, also show differences: in the interior part
  (${\la}10^4$~km~s$^{-1}$), the variations are mostly due to the different
  hydrodynamic evolution of the deflagration flame for the different ignition
  conditions, but in the outer part, the ejecta are shifted to lower velocities
  for lower C mass fractions.  Especially $^{56}$Ni and stable iron isotopes are
  confined to lower velocities; also the peak of the $^{28}$Si distribution
  shifts to lower velocities. Moreover, comparing the homogeneous models, the
  outer layers of the ejecta contain more unburnt material for a smaller C mass
  fraction (compared to the initial composition), which can be explained by less
  burning at these smaller densities for lower C mass fractions (see
  Appendix~\ref{sec:calibration}). The more realistic model with a C depleted
  core resembles the 50\% model in the outer layers (less unburnt material) and
  the 30\% model in the inner layers (the ejecta are shifted to lower
  velocities).

\subsection{3D Simulations}

  \begin{sidewaystable}[p]
  \centering
  \caption{Average initial C mass fraction, $^{56}$Ni mass, asymptotic kinetic
  energy, rise time to \textit{B}-band maximum, peak magnitudes, decline rate,
  and spectral velocities of the Si\,\textsc{ii} feature at $\lambda6355\,\AA{}$.
  }
  \label{tab:3dproperties}
  \begin{tabular}{lccccccccccc}
    \hline\hline
    Model & $\overline{X}_0(^{12}\mathrm{C})$ & $M(^{56}\mathrm{Ni})$ &
    $E_\mathrm{kin,a}$ & 
    $t(B_\mathrm{max})$ & 
    $U_{\mathrm{max}}$ & 
    $B_{\mathrm{max}}$ & 
    $V_{\mathrm{max}}$ & 
    $R_{\mathrm{max}}$ & 
    $I_{\mathrm{max}}$ & 
    $\Delta m_{15}(B)$ &
    $v_\mathrm{Si}(t_{B_\mathrm{max}})$ \\
    &  & $(M_\odot)$ &
    ($10^{51}$ erg) & 
    (d) & 
    (mag) & 
    (mag) & 
    (mag) & 
    (mag) & 
    (mag) & 
    (mag) &
    $(10^3\ \mathrm{km}\ \mathrm{s}^{-1})$ \\
    \hline
    rpc20\_DDT8\_N100             & 0.26 & 0.364 & 0.43 & --- &  --- & --- & --- & --- & --- & --- & --- \\
    rpc32\_DDT8\_N100             & 0.36 & 0.603 & 1.28 & 17.4 & $-18.9$ & $-19.0$ & $-19.5$ & $-19.5$ & $-19.5$ & 1.41 & 12.5 \\
    rpc40\_DDT8\_N100             & 0.42 & 0.701 & 1.43 & 17.5 & $-19.1$ & $-19.2$ & $-19.7$ & $-19.6$ & $-19.5$ & 1.49 & 13.5 \\
    c50\_DDT8\_N100               & 0.50 & 0.799 & 1.54 & 17.3 & $-19.4$ & $-19.4$ & $-19.9$ & $-19.7$ & $-19.6$ & 1.55 & 14.4 \\
    N100 \citep{seitenzahl2013a}  & 0.50 & 0.604 & 1.44 & 16.6 & $-18.8$ & $-19.0$ & $-19.5$ & $-19.6$ & $-19.6$ & 1.42 & 13.1 \\
    \hline
  \end{tabular} 
  \end{sidewaystable}

  A few 3D full-star simulations with more realistic treatment of
  the 3D turbulent burning process have been conducted to quantify the impact of
  different compositions.  In all simulations, the initial model is composed of
  a core of differing C mass fraction ($0.2$ to $0.5$) which is surrounded by an
  outer layer with a C mass fraction of $0.5$. General properties of the model
  and results from the radiative transfer simulations are presented in
  Table~\ref{tab:3dproperties}. The detailed nucleosynthetic abundances for
  selected models are given in Table~\ref{tab:ryields} for radioactive isotopes
  after 100~s (this corresponds to the end of the simulation, when the expansion
  is nearly homologous) and in Table~\ref{tab:syields} for stable isotopes after
  2~Gyr.  For these models, the $^{56}$Ni mass varies between 0.36 and 0.80
  solar masses.  The relative abundances (normalized to $^{56}$Fe) compared to
  solar values do not show large changes for different composition; here,
  metallicity (i.e., mainly the neutron-rich isotope $^{22}$Ne after He burning)
  plays a larger role \citep[see][Fig.~7]{seitenzahl2013a}.

\begin{figure}
    \centering
    \includegraphics{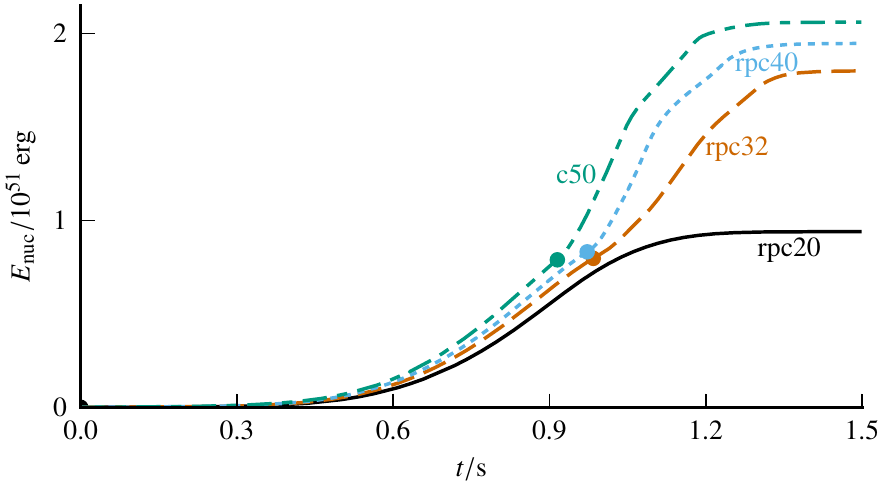}
    \caption{Evolution of nuclear energy in time for several 3D simulations
      with different initial compositions but otherwise identical explosion
      parameters (models rpc20\_DDT8\_N100, rpc32\_DDT8\_N100,
      rpc40\_DDT8\_N100, c50\_DDT8\_N100; inner core of 20\%, 32\%, 40\% and
      50\% C by mass, respectively).  The dot marks the point of the first DDT\@.
      The rpc20 model failed to initiate a detonation for this DDT criterion.
    }
    \label{fig:3dtimeevolution}
  \end{figure}

  When the different parameters (ignition conditions, DDT criteria,
  and initial compositions) are varied, the hydrodynamic evolution in the 3D models shows
  similarities to the 2D models. This can be seen for different initial
  compositions, e.g., in the evolution of the nuclear energy, which is shown in
  Fig.~\ref{fig:3dtimeevolution} for selected 3D models. The main difference
  between the evolution in 2D and 3D models is the slower energy release in 3D
  which is due to burning starting from spheres unlike tori in
  2D-axisymmetric geometry. Hence, the deflagration transitions later to the
  turbulent regime driven by the Rayleigh--Taylor instability.

  In Fig.~\ref{fig:3dtimeevolution}, a dot indicates the time when the first
  DDT is initiated and thus marks the transition to the detonation phase. Model
  rpc20\_DDT8\_N100 fails to initiate a detonation because the released nuclear
  energy does not generate enough turbulent motions to trigger the
  DDT.\footnote{The DDT criterion chosen here requires high turbulent velocity
    fluctuations $\ge 10^8$\ cm\ s$^{-1}$ to be present with a certain
    probability for at least half an eddy-turnover-time; this is the same
    criterion as used by \citet{seitenzahl2013a}. More details on the treatment
    of the criterion are described by \citet{ciaraldi2013a}.} As this
  model is simply a pure deflagration (also called ``failed detonation'' by
  \citealp{jordan2012b,fink2014a}, see there for recent models), it will not be
  discussed further. Nevertheless it is interesting that the DDT criterion
  chosen here fails for some models, while it successfully initiates a
  detonation for other models.  For the model series with varying C fraction,
  the nuclear energy release during the deflagration phase rises with the
  carbon mass fraction as expected from the binding energy differences and from
  the calibration (see Appendix~\ref{sec:calibration}).  This leads to the DDT
  criterion being fulfilled earlier for larger C mass fractions; thus, the
  expansion of the WD is smaller and unburnt material is present at higher
  densities. The detonation consumes the remaining unburnt material; and for
  higher densities, more $^{56}$Ni is produced.  Moreover, for larger C mass
  fractions, the transition density to burning to NSE is smaller, adding to the
  effect of producing more NSE material for larger C mass fractions and thereby
  releasing more nuclear energy. The $^{56}$Ni mass follows a similar linear
  relation for the carbon mass fraction with a slope near 1, similar to the
  mean relation found for the 2D models (cf.\ Equation~\ref{eq:nimasslinear}
  and values in Tables~\ref{tab:3dproperties} and \ref{tab:ryields}).

  \begin{figure}
    \centering
    \includegraphics{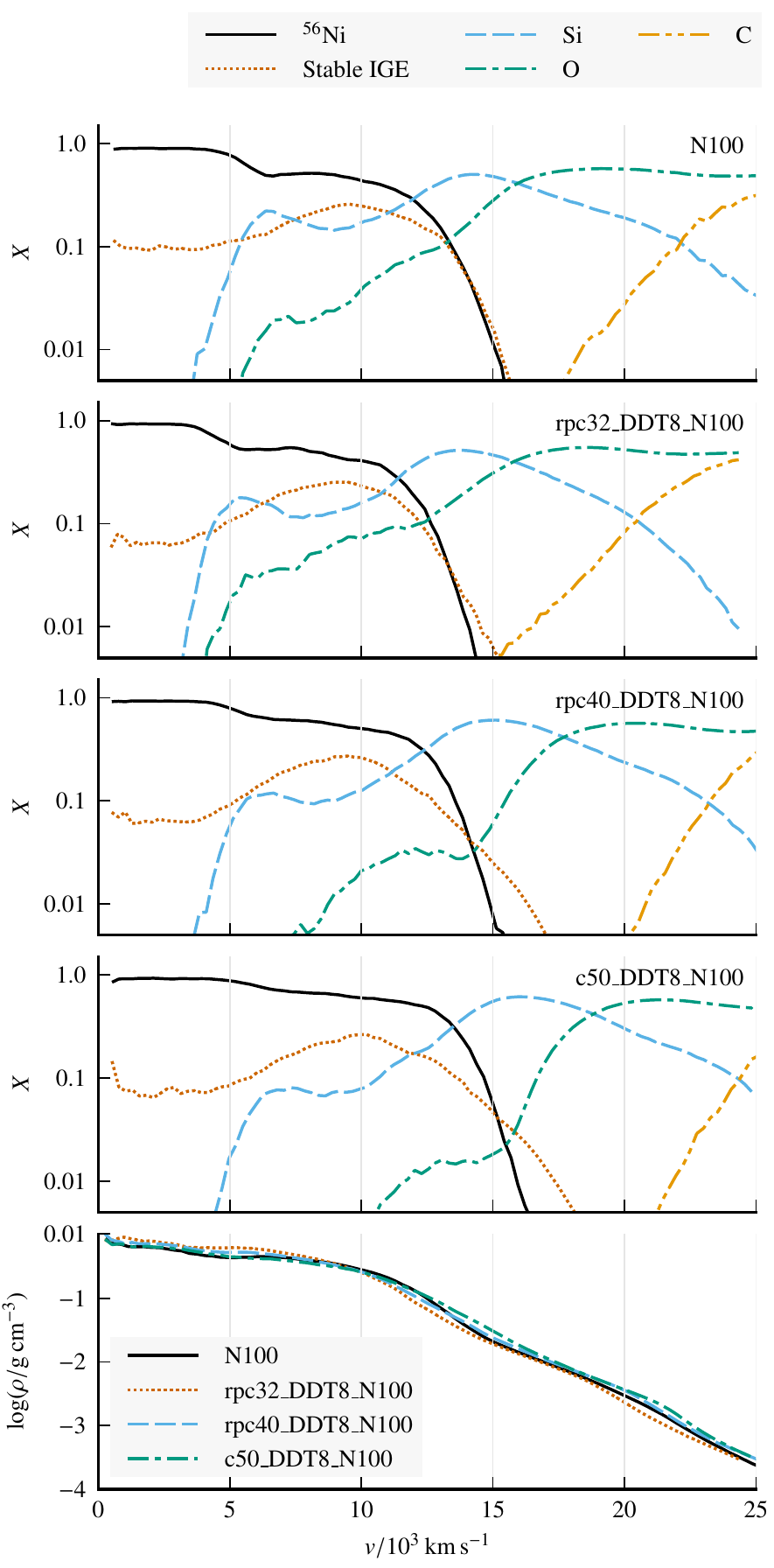}
    \caption{Angle averaged ejecta distribution in velocity space for several
      3D models.  The model in the top panel is N100 from
      \citet{seitenzahl2013a}. Shown are the mass fractions (top panels, stable
      IGE are all iron group elements with $Z>20$ without $^{56}$Ni) and the
      density (bottom panel).}
    \label{fig:3dejecta}
  \end{figure}
  \begin{table}
    \caption{Nucleosynthetic yields of radioactive isotopes (in solar masses)
      after $100$~s for different DDT8\_N100 models.
    }
    \label{tab:ryields}
    \centering
    \tiny
    \begin{tabular}{cccccccc}
      \hline\hline
      \input{tables/ryields}
      \hline
    \end{tabular}
  \end{table}
  \begin{table}
    \caption{Asymptotic nucleosynthetic yields of stable isotopes (in solar
      masses) for different DDT8\_N100 models.
    }
    \label{tab:syields}
    \centering
    \scriptsize
    \begin{tabular}{cccccccc}
      \hline\hline
      \input{tables/syields}
      \hline
    \end{tabular}
  \end{table}

  Apart from examining the influence of each parameter separately, several
  parameters can be changed at once to create models with similar $^{56}$Ni
  yields. As these models will show similar peak luminosities (dominated by the
  total amount of $^{56}$Ni), they can be compared in their ability to explain
  the same SN, as opposed to explaining SNe~Ia in general.  The spherically
  averaged ejecta distribution in velocity space is compared for two 3D models
  with similar $^{56}$Ni masses of about $0.6 M_\odot$ in
  Fig.~\ref{fig:3dejecta}: the first panel shows the homogeneous model (in
  initial composition) N100 from \citealp{seitenzahl2013a}, in the second panel the C
  depleted model rpc32\_DDT8\_N100 is plotted. The C depleted model features a
  lower asymptotic kinetic energy of $1.28\times 10^{51}$~erg compared to the
  homogeneous model ($1.44\times 10^{51}$~erg). The global structure of the
  ejecta is similar to the 2D models (Figs.~\ref{fig:velc30},
  \ref{fig:velrpc32}, \ref{fig:velc50}): outer layers of C, O and Si and a core
  consisting mainly of $^{56}$Ni and stable IGE. The stable iron group elements
  are created mainly in the deflagration ashes during normal freeze-out from
  NSE\@.  This is also the reason for the stable iron group elements extending
  to rather high velocities for both models, up to
  $\sim15\times10^3$~km~s$^{-1}$. They are created during the deflagration phase
  in the rising hot plumes, thus being present at large radii and
  velocities\footnote{This is a general feature of multi-dimensional DDT models;
    as opposed to 1D models, where the stable iron group elements
    are concentrated near the center (see fig.\ 2 from
    \citealp{khokhlov1991b}).}.  Despite the larger kinetic energy in the
  homogeneous model, the velocities in the ejecta tend to be only slightly
  larger than in the carbon depleted model with similar $^{56}$Ni mass (see
  first and second panel of Fig.~\ref{fig:3dejecta}).  Especially the outer
  boundary of the Ni core and the maximum of the Si distribution are shifted
  by only a few $100$~km~s$^{-1}$. Moreover, more unburnt material is present
  in the C depleted model mostly because of the shift in the burning tables (see
  Appendix~\ref{sec:calibration}).  When comparing a model series with varying
  core C mass fraction (second, third, and fourth panel of
  Fig.~\ref{fig:3dejecta}), these effects can be seen more clearly as the
  kinetic energy of the ejecta increases with increasing C mass fraction and
  increasing production of $^{56}$Ni.

  The density structure (bottom panel of Fig.~\ref{fig:3dejecta})
  is very similar for all models; thus, the differences in the spectra 
  mainly stem from differences in the abundance distributions.

\section{Synthetic Observables}
\label{sec:resultsrt}

  In this section, we present synthetic light curves and spectra from the
  radiative transfer simulations and compare to observed SNe. The effects of the
  initial composition are examined in two ways:
  \begin{enumerate}
    \item by comparing a series of models with differing carbon mass fraction
      but otherwise identical explosion parameters, thus having different
      kinetic energies and $^{56}$Ni masses;
    \item by comparing models with different carbon mass fraction producing a
      similar amount of $^{56}$Ni while having different kinetic energies.
  \end{enumerate}

\subsection{Light Curves}

  Radiative transfer simulations were run for the DDT models compared in
  Fig.~\ref{fig:3dtimeevolution} (rpc32\_DDT8\_N100, rpc40\_DDT8\_N100, and
  c50\_DDT8\_N100). The ignition condition was chosen to be the same as for the
  N100 model of \citet{seitenzahl2013a} since the intermediate deflagration
  strength of this model leads to the best agreement with observed light curves
  and spectra in this series \citep{sim2013a}\footnote{This model was also
    compared to SN~2011fe alongside a double-degenerate violent merger model in
    \citet{roepke2012a}.}.
  Compared to the N100 model, the DDT criterion of the new models was adjusted
  such that the rpc32 model produces approximately $0.6 M_\odot$ of
  $^{56}$Ni, the same amount as the N100 model of \cite{seitenzahl2013a}. Thus,
  these two models are very similar (apart from the initial composition) in
  order to assess the results of changing the initial C fraction.

  \begin{figure*}
    \centering
    \includegraphics{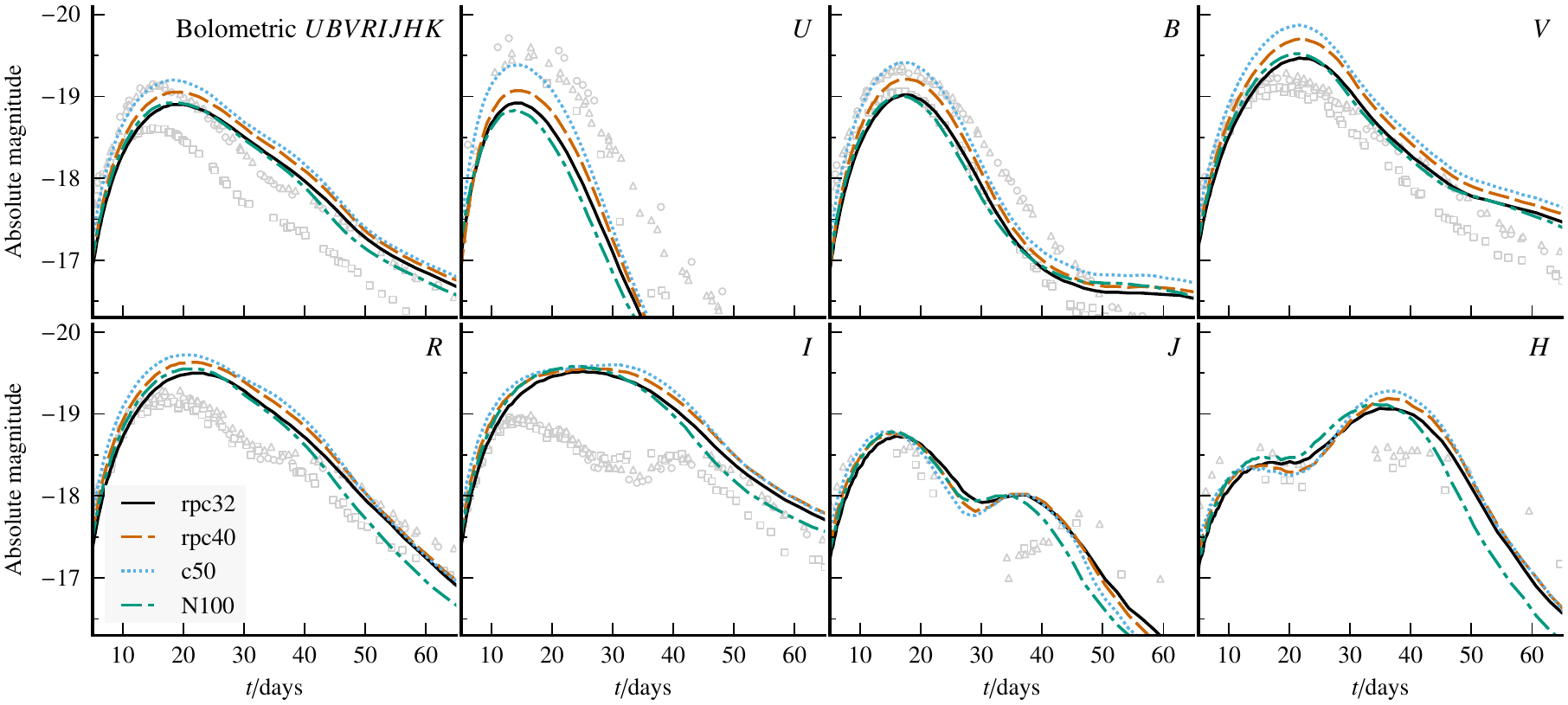}
    \caption{Angle-averaged light curves for models rpc32\_DDT8\_N100,
      rpc40\_DDT8\_N100, c50\_DDT8\_N100, and N100 of \citet{seitenzahl2013a}
      (solid lines) compared to some normal SNe~Ia (gray points; circles: SN~2003du,
      squares: SN~2004eo, triangles: SN~2005cf).
      }
    \label{fig:lightcurves}
  \end{figure*}
  
  The synthetic light curves from these models and the N100 model
  from \citet{seitenzahl2013a} are compared to some normal SNe~Ia in
  Fig.~\ref{fig:lightcurves}. The peak luminosity of the bolometric and the
  band-limited light curves are larger for larger $^{56}$Ni masses. The spread
  in peak luminosities is largest in the \textit{U} band and decreases to redder
  bands, similar to what is found by \citet{sim2013a} for their model series.

  The shapes of the light curves match observations quite well around maximum for
  \textit{U}, \textit{B}, and \textit{V} bands, although the flux is too low in
  the \textit{U} band and too high in the \textit{V} band. Thus, the colors are
  too red compared to observed light curves (similar to the models of
  \citealp{sim2013a}).  As already stated in \citet{sim2013a}, this is probably
  due to line-blocking mainly of IGE in the blue part of the spectrum. This is a
  generic feature of their spontaneous DDT models caused by the deflagration ashes
  (which contain stable IGE) rising to rather high velocities, near the IME
  (as described above, see Fig.~\ref{fig:3dejecta}), hence influencing the
  synthetic observables in the photospheric phase. In contrast to this, in 1D
  models (see fig.\ 2 from \citealp{khokhlov1991b}) the IGE are contained in
  the core of the ejecta beneath the radioactive $^{56}$Ni owing to the
  spherical symmetry adopted in these models, neglecting the turbulent
  deflagration burning. Apart from this, the reddening could also be due to
  shortcomings in the radiative transfer treatment, as reproducing the colors
  in radiative transfer simulations of SNe~Ia is in general difficult
  \citep{dessart2013b}. 

  In the \textit{I} band, the models deviate from observed light curves: they
  are too bright and do not show two maxima, similar to the models in
  \citet{sim2013a}. Although this may be due to an incomplete treatment in the
  radiative transfer code affecting the Ca\,\textsc{ii} infrared triplet, which
  significantly contributes to this band \citep{sim2013a}, this may also hint to
  the spontaneous DDT models being in general inferior to other models in this
  respect. For example, sub-Chandrasekhar models \citep{sim2010a} or violent
  merger models \citep{pakmor2012a} show better agreement using the same
  radiative transfer code \textsc{Artis} (see also \citealp{sim2013a}).
  
  In the near-infrared bands \textit{J} and \textit{H}, the models agree
  qualitatively with observations, matching the magnitudes at the first maximum
  and exhibiting a second maximum. The variations in these near-infrared bands
  are, especially at maximum, smaller than in the optical bands, which is
  also seen in observations showing that SNe~Ia are better standard candles in
  the near-infrared \citep{elias1985a,meikle2000a,krisciunas2004a}.
  Moreover, in these bands, the first maxima are larger compared to the light
  curves in \citet{sim2013a}, thus agreeing better with observed light curves.  As
  already predicted in \citet{sim2013a}, this results from using a larger atomic
  data set, thus producing more fluorescence in the near-infrared.
  The position of the second maximum, however, is too early compared to observed
  light curves. The second maximum is caused by the recombination front from
  doubly to singly ionized material hitting the iron-rich core
  \citep{kasen2006b}.  Thus, the offset between simulations and observations
  could indicate that IGE reside at too large velocities in our models.
  However, it could also be related to  deficiencies in the numerical
  treatment, such as inaccurate atomic data or approximations in calculating
  the plasma state in \textsc{Artis}.

  A comparison of the two models with similar $^{56}$Ni masses (N100 from
  \citealp{seitenzahl2013a} and rpc32\_DDT8\_N100) shows only slight differences.
  The main consequence on the light curve here is given by the different kinetic
  energies of the ejecta. According to the analytic study of bolometric light
  curve models by \citet{pinto2000a}, models with larger kinetic energy ``peak
  earlier, at higher luminosities, and decline more rapidly'' \citep[][see also
  their fig.~4]{pinto2000a}.  This is indeed also found for the bolometric
  light curves of the models N100 from \citet{seitenzahl2013a} and
  rpc32\_DDT8\_N100 (see Fig.~\ref{fig:lightcurves}): the C depleted model
  peaks later and at a lower luminosity. Moreover, its decline rate is smaller.
  The effect is not as large as for the models in \citet{pinto2000a} because the
  total kinetic energy of the rpc32\_DDT8\_N100 model differs only by about 11\%
  from the N100 model of \cite{seitenzahl2013a}.

\subsection{Width--Luminosity Relation}

  \begin{figure}
    \centering
    \includegraphics{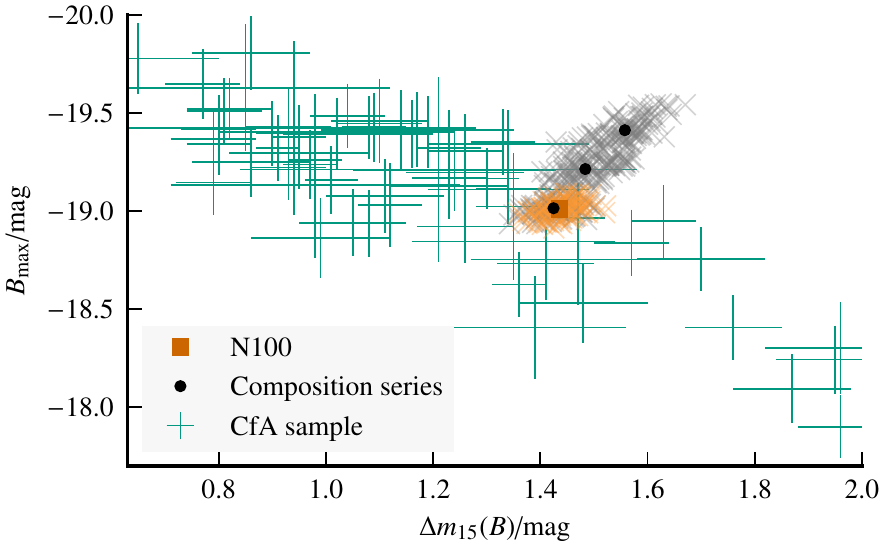}
    \caption{Light curve width--luminosity relation for a series of models with
    differing C depletion in the core, but otherwise identical parameters
    (in black: models rpc32\_DDT8\_N100, rpc40\_DDT8\_N100, and
    c50\_DDT8\_N100 with increasing \emph{B}$_\mathrm{max}$ in this order) as
    well as for the N100 model of \citet[in gold]{seitenzahl2013a}.  The dots
    and the square denote angle-averaged values, pale crosses denote values for
    different lines of sight.  The green crosses show observed supernovae from
    the CfA sample \citep{hicken2009b}.
    }
    \label{fig:dm15}
  \end{figure}

  The decline in the \textit{B} band of the models is more rapid than for most
  normal SNe~Ia (Fig.~\ref{fig:dm15}). More importantly, the model series as a
  whole fails to show the same width--luminosity relation (WLR) as normal SNe~Ia;
  in contrast, the WLR is roughly perpendicular to that observed
  (Fig.~\ref{fig:dm15}). The fundamental parameters for the light curve
  evolution that are changed in this model series are the kinetic energy of the
  ejecta and the $^{56}$Ni mass, which both increase with increasing C fraction.
  According to the analytic study of light curves by \citet{pinto2000a}, both of
  these parameters anti-correlate individually with the observed WLR\@. Therefore,
  it is not surprising to find an anti-correlation for our model series, where
  the increase in C fraction (as a physical parameter of the explosion model)
  leads to an increase in kinetic energy and $^{56}$Ni mass, both driving a
  trend perpendicular to the observed WLR\@. This implies that the initial
  composition is probably not the main parameter driving the WLR, but rather a
  secondary parameter causing scatter perpendicular to the WLR\@. This is similar
  to orientation effects also driving scatter around the mean WLR\@.

  The only possibility of driving the WLR in the direction that is observed 
  would, in this model, be the existence of a correlation of the physical model
  parameters. In this case, the ignition configuration and the DDT criterion
  would depend on the initial composition (in a yet unknown way), thereby 
  supposedly resulting in a suitable WLR\@. The 1D delayed detonation models
  of \citet{hoeflich1996a} show a WLR, where the changing parameter is the DDT
  transition density, but as this parametrization does not include turbulence,
  for example, it cannot be easily generalized to our multi-dimensional models.
  The 2D models of \citet{kasen2009a} lie in a reasonable region of the light
  curve width--luminosity diagram; this was reached by varying the ignition
  conditions for the deflagration as well as the DDT criterion. This model
  series faces the problem that the correlation between the varying explosion
  parameters and the underlying physical parameters of the initial model (such
  as central density, composition or metallicity) is not known and thus does
  not identify the physical parameter driving the WLR\@.
  
\subsection{Spectra}

\begin{figure}
    \centering
    \includegraphics{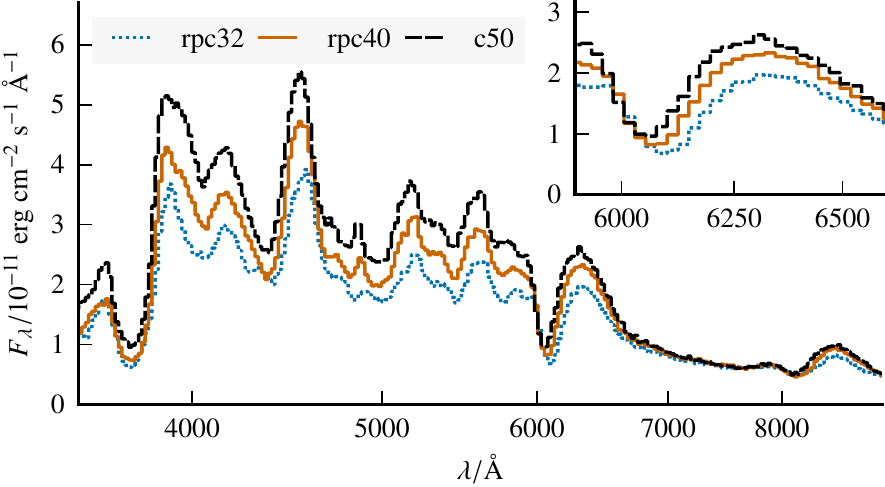}
    \caption{Comparison of spectra around \textit{B}-band maximum at 17.3 days
    post explosion for models rpc32\_DDT8\_N100, rpc40\_DDT8\_N100, and
    c50\_DDT8\_N100. The inset shows the Si~\textsc{ii} feature at
    $\lambda6355\,\AA{}$ in more detail, the units are the same as in the main
    plot. All fluxes are scaled to a distance of 1~Mpc.
    }
    \label{fig:spectra}
  \end{figure}

  The synthetic spectra of the model series are shown in Fig.~\ref{fig:spectra} at
  \textit{B}-band maximum. They share all main spectral features and differ mostly in
  the absolute flux values (Fig.~\ref{fig:spectra}). Moreover, the
  Si\,\textsc{ii} feature at $\lambda6355\,\AA{}$ varies in blue shift for
  different models: with increasing C mass fraction, the absorption feature
  shifts from $12.5\times10^3$~km~s$^{-1}$ (rpc32) to
  $14.4\times10^3$~km~s$^{-1}$ (c50), thus reflecting the change in the
  velocity distributions (compare Fig.~\ref{fig:3dejecta}). The features
  associated with Ca\,\textsc{ii}\footnote{The Ca\,\textsc{ii} H \& K lines
  ($\lambda\lambda3934\,\&\,3968$\,\AA{}) and the Ca\,\textsc{ii} infrared triplet.},
  however, are not shifted in wavelength for different models.
  
  \begin{figure}
    \centering
    \includegraphics{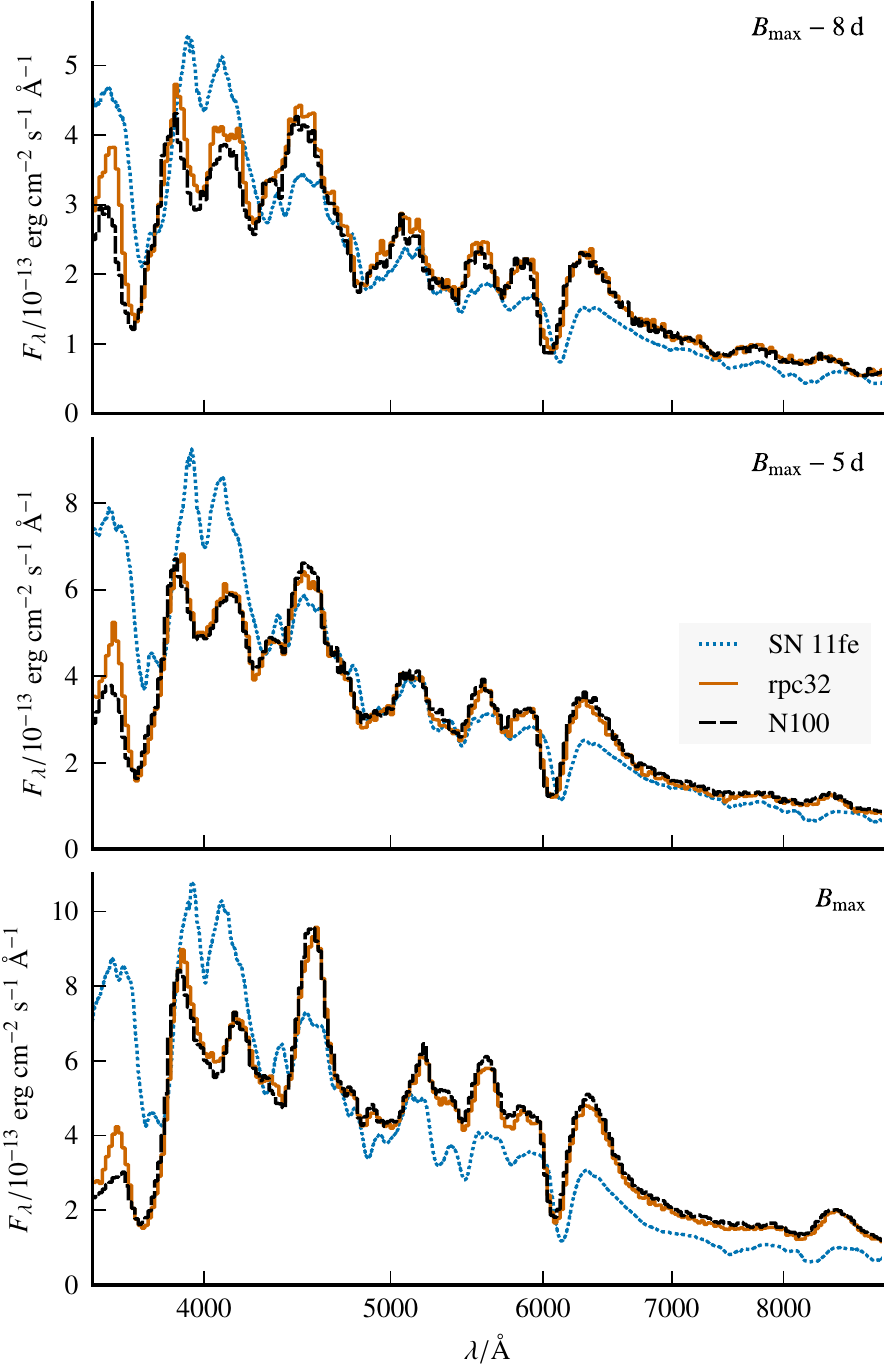}
    \caption{Comparison of spectra for three epochs (8~d before, 5~d before and
      at \textit{B}-band maximum) of models rpc32\_DDT8\_N100, and N100 of
      \citet{seitenzahl2013a} to SN~2011fe. The flux is scaled to M101, the
      host galaxy of SN~2011fe.
    }
    \label{fig:spectraseries}
  \end{figure}

  A comparison between two models with similar $^{56}$Ni mass but different
  kinetic energies (rpc32\_DDT8\_N100 and N100 from \citealp{seitenzahl2013a},
  see Fig.~\ref{fig:spectraseries}) shows that the C depleted model shifts in
  velocities only by about 600~km~s$^{-1}$ at \textit{B}-band maximum beacuse
  of the lower kinetic energy of the ejecta. Comparing to observations, this
  shift goes in the right direction but is not large enough to account for the
  lower velocities in, e.g., SN~2011fe, as shown for several epochs in
  Fig.~\ref{fig:spectraseries}.  Moreover, the bulk of observed SNe shows
  considerably lower velocities, mostly between $10\,000$~km~s$^{-1}$ and
  $12\,000$~km~s$^{-1}$ at \textit{B}-band maximum
  \citep{benetti2005a,silverman2012b}.  The magnitude of this effect can be
  estimated by assuming that the velocities in the two models scale with the
  square root of the kinetic energy, $\frac{\overline{v}_1}{\overline{v}_2} =
  \sqrt{\frac{E_{\mathrm{kin},1}}{E_{\mathrm{kin},2}}}$. A change in the
  kinetic energies from N100 to the rpc32 model of about 11\% hence results in
  a change in velocities of about 6\%, which yields about 750~km~s$^{-1}$ for
  the Si\,\textsc{ii} velocity, on the order of the change seen in the models. 

  This seems to be a shortcoming of the spontaneous DDT model, also present in
  our previous model series \citep{seitenzahl2013a,sim2013a}, but also in
  studies of other groups (e.g., the DDC models from \citealp{dessart2013a},
  see their fig.~17; these models are based on explosion simulations of
  \citealp{khokhlov1991a}). 
  
  Thus, if explosions from Chandrasekhar-mass WDs constitute a large fraction
  of normal SNe Ia, the nucleosynthetic yields (and thus the energy release)
  could be affected by uncertainties in the nuclear reaction rates.
  Alternatively, the mechanism that distorts the hydrostatic equilibrium of the
  WD\footnote{This is needed for the detonation to propagate through
  lower-density, pre-expanded material such that also IME are produced.} may
  not involve a direct transition from the deflagration to a detonation, as in
  our spontaneous DDT models, but perhaps other large-scale motions (e.g.,
  pulsations).

\subsection{Comparison to Earlier Studies}

  \citet{hoeflich1998a} presented a series of 1D delayed detonation models
  (including hydrodynamics, nucleosynthesis, light curves, and spectra) with
  varying metallicity and included one model with a different C fraction.
  Their 1D models treat the deflagration as propagating at a certain fraction
  of the local sound speed; the transition to a detonation is initiated when
  a certain density is reached. Despite these simplifications, their conclusions
  are similar to our findings: a lower C fraction leads to lower $^{56}$Ni
  production, less kinetic energy, and more confined ejecta. Their C
  reduced model shows a faster decline, as opposed to what is seen in our
  models and expected for lower kinetic energies according to the analytic
  study of \citep{pinto2000a}. They also mention that more realistic structures
  of the progenitor composition should be taken into account including an outer
  accretion layer and a C depleted core; this was accomplished in the present
  study. Their assumption, however,  that a model with a homogeneous, but lower
  C fraction does not show a difference to a model with a C depleted core (see
  their ``Final Discussion and Conclusions'') holds only to a first
  approximation. As the propagation of the burning fronts does not only depend
  on density, but also on the composition (see Appendix~\ref{sec:calibration}),
  the energy release and burning products of the detonation in the outer layers
  depend on the composition there.

  The 1D simulations by \citet{umeda1999a} include only hydrodynamics and
  nucleosynthesis of delayed detonation models, where the DDT density depends on
  the initial C fraction. In contrast to this, our DDT criterion includes
  different effects as, e.g., turbulent velocity fluctuations (see
  Section~\ref{sec:leafs}). Nevertheless, this leads to a similar result: the
  density at which the detonation is initiated decreases for decreasing C
  fractions. Thus, in their model series, the produced amount of $^{56}$Ni
  decreases for decreasing C fractions (see their fig.~2). Apart from this,
  the working hypothesis of \citet{umeda1999a}, who assume the C mass fraction to
  be responsible for the WLR, is challenged by our findings.  

  \new{
  In their study of 1D delayed detonation models, \citet{dominguez2001a} compute
  stellar models for different ZAMS masses and metallicities which they use as
  initial models for the explosion simulations. All models use the same central
  density at explosion and employ the same density as DDT criterion, but do not
  take the pre-explosion simmering phase into account. They find that for larger
  ZAMS masses, less $^{56}$Ni is produced because of the lower C abundance,
  resulting in lower velocities in the ejecta, similar to our findings. This
  leads in their models to a decrease in the maximum brightness, while the
  decline rate stays constant. Thus, \citet{dominguez2001a} conclude that the
  variation in the ZAMS mass leads to a spread or offset in the WLR, similar to
  what we find for the initial C mass fraction.
  }

  \citet{hoeflich2010a} suggest that in addition to the primary light curve
  parameter, $\Delta m_{15}$ or stretch $s$, two independent parameters are
  necessary to describe the differences in shapes for different SNe. As physical
  parameters they suggest C/O ratio and central density to account for different
  shapes in the early and late phase of the light curve, respectively. In the
  1D models of \citet{hoeflich2010a}, the transition density of the DDT
  determines the $^{56}$Ni mass of the explosion; the impact of other
  parameters (C/O ratio, central density) on the intrinsic brightness is small;
  nevertheless, these variations should be taken into account in the
  calibration. Our study agrees on the C fraction being a secondary parameter in
  the family of SN~Ia light curves. In our 3D models, however, the C fraction
  causes large variations in the $^{56}$Ni mass because of the different turbulent
  evolution of the deflagration flame and the resulting different triggering of
  the DDT\@. This should be taken into account when trying to create a physically
  motivated multi-parameter set for SN~Ia light curves.

  The first multi-dimensional simulations examining different C fractions were
  presented by \citet{roepke2004c} and \citet{roepke2006b}. In their 3D
  simulations of pure deflagrations, the C fraction does not affect the
  explosion significantly; only the kinetic energy of the ejecta is altered to some
  extent. Therefore, they conclude that ``the progenitor's carbon-to-oxygen
  ratio is unlikely to account for the observed variations in type Ia supernova
  luminosity'' \citep{roepke2004c}. This statement only holds for pure
  deflagration models, which nowadays are thought to account rather for
  2002cx-like SNe~Ia than for normal SNe~Ia
  \citep{jordan2012a,kromer2013a,fink2014a}
  because of their mixed ejecta structure in contrast to the layered structure
  seen in normal SNe~Ia. Thus, their statement does not apply to modeling normal
  SNe Ia and  it does not contradict our results for delayed
  detonation models.
  
\section{Conclusions}
\label{sec:conclusions}

  In this work, we study the hydrodynamics, nucleosynthesis, synthetic light
  curves, and synthetic spectra of a series of \new{multi-dimensional}
  spontaneous DDT models for SNe~Ia in order to \new{examine if varying the
  initial C fraction resolves remaining discrepancies to observations.
  The main points we consider are the WLR resulting from the models and
  differences in spectral features.
  }

  Firstly, the initial C mass fraction is not the primary parameter of SNe~Ia
  (at least for spontaneous DDT models). Although absolute luminosities
  \new{(\textit{B}$_\mathrm{max}$ between $-19.0$ and $-19.4$)} and decline
  rates \new{($\Delta m_{15}(B)$ between $1.41$ and $1.55$)} are in the range
  of normal SNe~Ia, respectively, our model series fails to reproduce the
  observed WLR (Fig.~\ref{fig:dm15}).  Therefore, it is probably only a
  secondary parameter causing scatter perpendicular to the observed WLR\@. This
  may only be changed by a concerted correlation of the different physical
  parameters of the underlying explosion model, such as ignition conditions or
  DDT criteria.

  Secondly, carbon depleted models do not show significantly better agreement of
  important spectral features, such as the Si\,\textsc{ii} feature at
  $\lambda6355\,\AA{}$. The decrease in kinetic energy does not lead to a
  decrease in the blueshift of the feature to be compatible with 
  the bulk of normal SNe~Ia. This shortcoming seems to be generally present in
  spontaneous DDT models (1D, 3D, different groups; see discussion above).

  \new{Finally, our spontaneous DDT models are able to reproduce most of
  the observed properties of SNe Ia light curves and spectra, thus supporting
  the spontaneous DDT model. So far, however, our 3D spontaneous DDT models
  do not show the observed width--luminosity relation.}
  While the deflagration strength (through number of ignition kernels,
  \citealp{seitenzahl2013a,sim2013a}) and the initial C fraction (this work) are
  not the primary parameter, it may still be possible that other parameters
  (e.g., DDT criterion) or yet unknown correlations of parameters are able to
  reproduce the light curve width--luminosity relation in 3D models.
  Nevertheless, other shortcomings remain, such as colors, which are too red
  \citep{sim2013a}, and the velocities of spectral features, especially the
  Si\,\textsc{ii} feature that is defining SNe~Ia. This may be interpreted in
  different ways: if Chandrasekhar-mass progenitors are indeed responsible for
  the bulk of SNe~Ia, the spontaneous DDT model has some severe shortcomings;
  this may hint to the possibility that a different mechanism distorts the
  hydrostatic equilibrium of the WD and leads to a detonation (e.g.,
  pulsations). Apart from this, the failure of recent multi-dimensional DDT
  models to identify the primary parameter of the WLR could also indicate that
  this primary parameter is the mass of the primary
  WD\footnote{This was already suspected by \citet{pinto2000a} in their
    analytic study of light curves and is supported by observations of
    \citet{stritzinger2006a} and \citet{scalzo2014a}.}, 
  as is the case in detonations of sub-Chandrasekhar-mass WDs either in a
  double degenerate binary (violent merger scenario, e.g.,
  \citealp{pakmor2012a,pakmor2013a}) or in a single degenerate system (double
  detonation scenario, e.g., \citealp{fink2010a}).

  \begin{acknowledgements}
    The 3D models have been computed on the supercomputers
    \textsc{Jugene} and \textsc{Juqueen} at the J\"{u}lich Supercomputer Center
    under the project HMU13.
    This work was also supported by the Deutsche
    Forschungsgemeinschaft via the Transregional Collaborative Research
    Center TRR 33 ``The Dark Universe'', the Emmy Noether Program (RO
    3676/1-1), the ARCHES prize of the German Ministry of Education and
    Research (BMBF), the graduate school ``Theoretical Astrophysics and
    Particle Physics'' at the University of W\"urzburg (GRK 1147) and the
    Excellence Cluster EXC~153.
    Parts of this research were conducted by the Australian Research Council
    Centre of Excellence for All-sky Astrophysics (CAASTRO), through project
    number CE110001020.
    STO acknowledges support from the Studienstiftung des deutschen Volkes and
    thanks S.~Hachinger for valuable dicussions. 
    RP acknowledges support  by the European Research Council under 
    ERC-StG grant EXAGAL-308037 and by the Klaus Tschira Foundation.
    MF, SAS, and FKR acknowledge travel support by the DAAD/Go8
    German-Australian exchange program.
    We thank S.~Taubenberger for providing the data of the CfA sample.
    For data processing and plotting, we used NumPy and SciPy
    \citep{oliphant2007a}, IPython \citep{perez2007a}, and Matplotlib
    \citep{hunter2007a}.
  \end{acknowledgements}

\bibliographystyle{aa} 
\bibliography{astrofritz}

\clearpage
\begin{appendix}

\section{Iterative Calibration of the Levelset Tables}
\label{sec:calibration}

  The tables necessary for determining the composition behind the burning fronts
  are created using an iterative calibration scheme similar to
  \citet{fink2010a}. This calibration scheme is carried out for homogeneous
  compositions of the progenitor WD ($X(^{12}\mathrm{C})=0.2,0.3,\ldots,0.9$),
  separately for deflagrations and detonations. It yields the composition behind
  the burning front as a function of the density of the unburnt material.

\begin{figure}[p]
    \centering
    \includegraphics{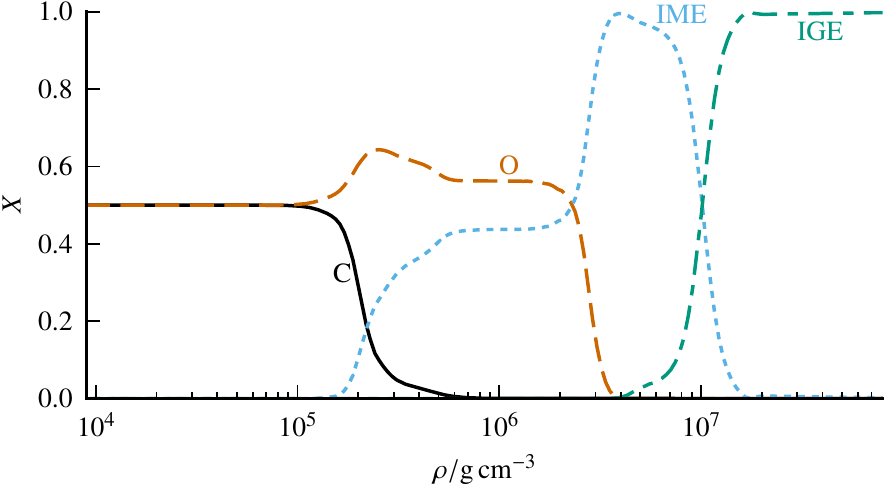}
    \caption{Levelset table for a detonation with an initial composition of
      $X(^{12}\mathrm{C})=0.5$. We show the mass fractions of C, O,
      intermediate mass elements (IME) and iron group elements (IGE), which are
      released behind the front, against the density of the unburnt material.
    }
    \label{fig:dettable-c50}
  \end{figure}

  Each calibration run uses as an initial estimate burning to NSE at the relevant
  densities (detonations: above $10^5$~g~cm$^{-3}$; deflagrations: above
  $2\times 10^5$~g~cm$^{-3}$), such that the energy release is overestimated.
  The table with the nucleosynthetic yields of this initial estimate as a function
  of density is used in a hydrodynamic simulation of a pure detonation or
  deflagration, followed by a nucleosynthetic postprocessing.  A new table is
  then computed with the detailed nucleosynthetic yields for use in the next
  hydrodynamic simulation. This procedure is iterated six times for each
  calibration run.  As an example, the final table for $X(^{12}\mathrm{C})=0.5$
  for detonations is shown in Fig.~\ref{fig:dettable-c50}. The transitions to
  different burning stages (C burning, O burning, Si burning) are clearly
  visible. The convergence of this scheme is based on the fact that the reaction
  rates depend strongly on density.  The overestimation of the energy release in
  the first hydrodynamic simulation is thus decreased by the subsequent
  postprocessing, since the density of unburned material prior to the front
  crossing is not strongly affected by a larger energy release.  

\begin{figure}[p]
    \centering
    \includegraphics{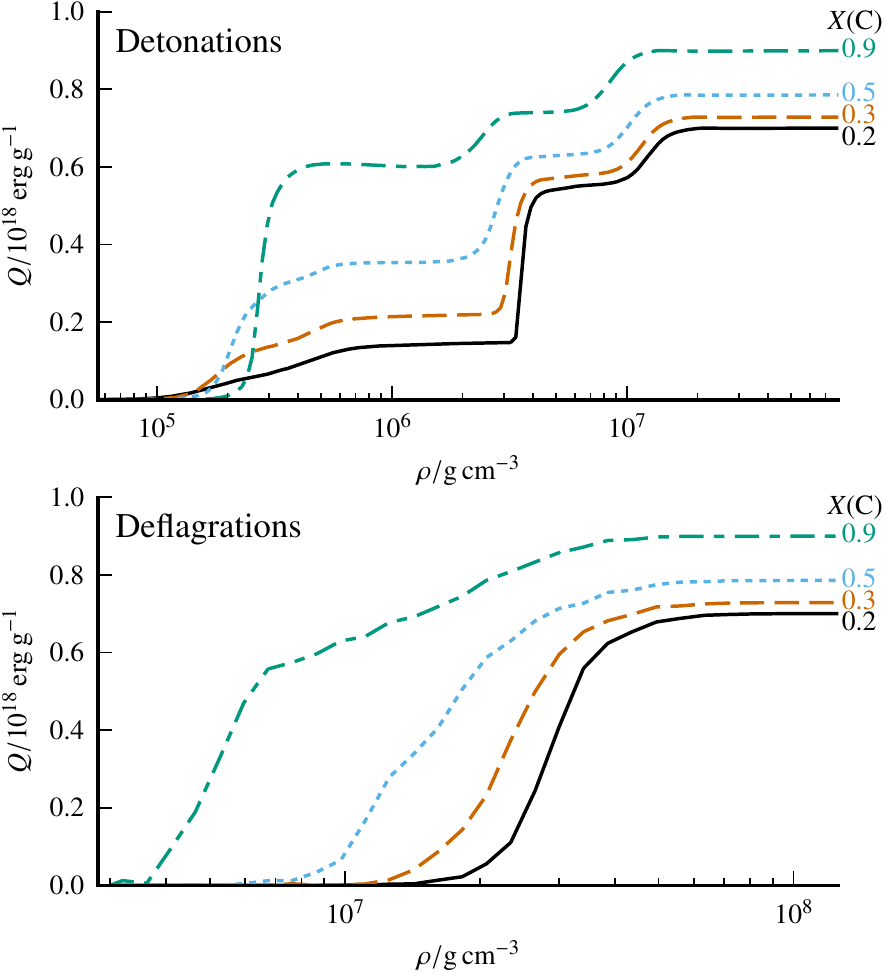}
    \caption{$Q$ value (energy release behind burning front) of levelset tables for selected initial
      compositions for detonations (upper panel) and deflagrations (lower panel).
    }
    \label{fig:tables-qvalue}
  \end{figure}

  The influence of the initial composition on the final tables is better seen by
  comparing the reaction $Q$ value, which is the energy release of the burning
  front. The $Q$ value is shown for different initial compositions in the upper
  panel of Fig.~\ref{fig:tables-qvalue} for detonations and in the lower panel
  of Fig.~\ref{fig:tables-qvalue} for deflagrations. The tables differ mainly in
  the density interval over which the composition changes; it is wider and
  extends to lower densities for detonations.  Apart from this, the shape is
  different, as in the tables for deflagrations the transitions to the different
  burning stages are not separated.  For both detonations and deflagrations, the
  $Q$ value is globally lower for lower carbon mass fractions. Moreover, the
  transitions to the different burning stages shift to higher densities for
  lower carbon mass fractions. Globally seen, this leads to a \textit{weaker
  development} of both detonations and deflagrations for lower carbon mass
  fractions.

  \end{appendix}

\end{document}

%% file: tables/ryields.tex
 & rpc32 & rpc40 & c50 \\
\hline
$^{14}$C  & 1.57E$-$06 & 3.62E$-$07 & 2.41E$-$07 \\
$^{22}$Na & 9.06E$-$09 & 2.01E$-$09 & 1.18E$-$09 \\
$^{26}$Al & 7.45E$-$07 & 2.02E$-$07 & 1.26E$-$07 \\
$^{32}$Si & 5.34E$-$09 & 1.34E$-$09 & 8.79E$-$10 \\
$^{32}$P  & 2.22E$-$07 & 1.05E$-$07 & 7.02E$-$08 \\
$^{33}$P  & 1.73E$-$07 & 8.07E$-$08 & 5.22E$-$08 \\
$^{35}$S  & 1.78E$-$07 & 6.11E$-$08 & 3.83E$-$08 \\
$^{36}$Cl & 3.50E$-$07 & 1.83E$-$07 & 1.24E$-$07 \\
$^{37}$Ar & 2.80E$-$05 & 2.04E$-$05 & 1.42E$-$05 \\
$^{39}$Ar & 4.82E$-$08 & 1.06E$-$08 & 6.76E$-$09 \\
$^{40}$K  & 4.73E$-$08 & 1.68E$-$08 & 1.06E$-$08 \\
$^{41}$Ca & 6.04E$-$06 & 4.03E$-$06 & 2.64E$-$06 \\
$^{44}$Ti & 1.18E$-$05 & 1.22E$-$05 & 1.18E$-$05 \\
$^{48}$V  & 9.31E$-$08 & 6.77E$-$08 & 5.28E$-$08 \\
$^{49}$V  & 4.56E$-$07 & 2.92E$-$07 & 2.29E$-$07 \\
$^{48}$Cr & 3.61E$-$04 & 3.80E$-$04 & 3.73E$-$04 \\
$^{49}$Cr & 2.36E$-$05 & 2.49E$-$05 & 2.33E$-$05 \\
$^{51}$Cr & 6.06E$-$06 & 5.58E$-$06 & 5.25E$-$06 \\
$^{51}$Mn & 8.61E$-$05 & 8.92E$-$05 & 8.15E$-$05 \\
$^{52}$Mn & 5.47E$-$06 & 5.12E$-$06 & 4.51E$-$06 \\
$^{53}$Mn & 2.14E$-$04 & 2.08E$-$04 & 2.00E$-$04 \\
$^{54}$Mn & 2.77E$-$06 & 2.83E$-$06 & 2.93E$-$06 \\
$^{52}$Fe & 9.11E$-$03 & 9.31E$-$03 & 9.00E$-$03 \\
$^{53}$Fe & 9.68E$-$04 & 9.73E$-$04 & 9.07E$-$04 \\
$^{55}$Fe & 1.88E$-$03 & 1.83E$-$03 & 1.76E$-$03 \\
$^{59}$Fe & 1.74E$-$06 & 2.77E$-$07 & 1.59E$-$07 \\
$^{60}$Fe & 9.91E$-$06 & 1.62E$-$06 & 8.55E$-$07 \\
$^{55}$Co & 1.10E$-$02 & 1.09E$-$02 & 1.01E$-$02 \\
$^{56}$Co & 1.27E$-$04 & 1.23E$-$04 & 1.13E$-$04 \\
$^{57}$Co & 9.06E$-$04 & 8.88E$-$04 & 8.58E$-$04 \\
$^{58}$Co & 4.27E$-$06 & 4.28E$-$06 & 4.38E$-$06 \\
$^{60}$Co & 4.18E$-$06 & 9.02E$-$07 & 4.54E$-$07 \\
$^{56}$Ni & 6.03E$-$01 & 7.01E$-$01 & 7.99E$-$01 \\
$^{57}$Ni & 1.53E$-$02 & 1.77E$-$02 & 2.03E$-$02 \\
$^{59}$Ni & 4.05E$-$04 & 4.08E$-$04 & 4.07E$-$04 \\
$^{63}$Ni & 3.26E$-$06 & 7.75E$-$07 & 4.80E$-$07 \\
$^{62}$Zn & 1.79E$-$04 & 3.04E$-$04 & 4.59E$-$04 \\
$^{65}$Zn & 1.39E$-$06 & 7.04E$-$07 & 5.03E$-$07 \\
$^{65}$Ga & 4.94E$-$08 & 7.95E$-$08 & 1.17E$-$07 \\
$^{68}$Ge & 4.14E$-$08 & 2.45E$-$08 & 1.86E$-$08 \\

%% file: tables/syields.tex
 & rpc32 & rpc40 & c50 \\
\hline
$^{12}$C  & 5.65E$-$03 & 1.50E$-$03 & 9.84E$-$04 \\
$^{13}$C  & 1.01E$-$08 & 3.59E$-$09 & 2.50E$-$09 \\
$^{14}$N  & 2.69E$-$06 & 5.94E$-$07 & 3.94E$-$07 \\
$^{15}$N  & 3.18E$-$09 & 6.57E$-$10 & 4.21E$-$10 \\
$^{16}$O  & 1.21E$-$01 & 5.52E$-$02 & 3.47E$-$02 \\
$^{17}$O  & 4.88E$-$07 & 1.14E$-$07 & 7.61E$-$08 \\
$^{18}$O  & 4.98E$-$09 & 8.15E$-$10 & 5.10E$-$10 \\
$^{19}$F  & 5.51E$-$10 & 6.70E$-$11 & 3.46E$-$11 \\
$^{20}$Ne & 6.44E$-$03 & 1.54E$-$03 & 8.98E$-$04 \\
$^{21}$Ne & 3.61E$-$07 & 7.18E$-$08 & 4.57E$-$08 \\
$^{22}$Ne & 4.79E$-$05 & 9.72E$-$06 & 6.14E$-$06 \\
$^{23}$Na & 5.10E$-$05 & 1.15E$-$05 & 7.13E$-$06 \\
$^{24}$Mg & 1.91E$-$02 & 9.08E$-$03 & 6.02E$-$03 \\
$^{25}$Mg & 6.59E$-$05 & 1.67E$-$05 & 1.02E$-$05 \\
$^{26}$Mg & 8.37E$-$05 & 1.98E$-$05 & 1.24E$-$05 \\
$^{27}$Al & 7.31E$-$04 & 2.94E$-$04 & 1.86E$-$04 \\
$^{28}$Si & 2.59E$-$01 & 2.48E$-$01 & 2.06E$-$01 \\
$^{29}$Si & 7.47E$-$04 & 3.77E$-$04 & 2.57E$-$04 \\
$^{30}$Si & 1.29E$-$03 & 6.19E$-$04 & 4.18E$-$04 \\
$^{31}$P  & 3.94E$-$04 & 2.11E$-$04 & 1.47E$-$04 \\
$^{32}$S  & 1.15E$-$01 & 1.14E$-$01 & 9.44E$-$02 \\
$^{33}$S  & 2.52E$-$04 & 1.67E$-$04 & 1.20E$-$04 \\
$^{34}$S  & 1.80E$-$03 & 1.21E$-$03 & 8.79E$-$04 \\
$^{36}$S  & 1.44E$-$07 & 4.31E$-$08 & 2.82E$-$08 \\
$^{35}$Cl & 1.27E$-$04 & 7.83E$-$05 & 5.44E$-$05 \\
$^{37}$Cl & 2.85E$-$05 & 2.06E$-$05 & 1.44E$-$05 \\
$^{36}$Ar & 2.04E$-$02 & 2.03E$-$02 & 1.73E$-$02 \\
$^{38}$Ar & 1.01E$-$03 & 6.98E$-$04 & 4.76E$-$04 \\
$^{40}$Ar & 5.52E$-$08 & 1.16E$-$08 & 7.30E$-$09 \\
$^{39}$K  & 9.47E$-$05 & 6.32E$-$05 & 4.14E$-$05 \\
$^{41}$K  & 6.07E$-$06 & 4.04E$-$06 & 2.64E$-$06 \\
$^{40}$Ca & 1.76E$-$02 & 1.76E$-$02 & 1.57E$-$02 \\
$^{42}$Ca & 2.72E$-$05 & 1.69E$-$05 & 1.04E$-$05 \\
$^{43}$Ca & 9.59E$-$08 & 3.45E$-$08 & 2.65E$-$08 \\
$^{44}$Ca & 1.19E$-$05 & 1.22E$-$05 & 1.18E$-$05 \\
$^{46}$Ca & 2.95E$-$08 & 5.29E$-$09 & 3.19E$-$09 \\
$^{48}$Ca & 2.93E$-$09 & 6.14E$-$10 & 3.97E$-$10 \\
$^{45}$Sc & 3.40E$-$07 & 1.96E$-$07 & 1.42E$-$07 \\
$^{46}$Ti & 1.49E$-$05 & 1.01E$-$05 & 6.67E$-$06 \\
$^{47}$Ti & 6.42E$-$07 & 4.60E$-$07 & 3.82E$-$07 \\
$^{48}$Ti & 3.62E$-$04 & 3.80E$-$04 & 3.73E$-$04 \\
$^{49}$Ti & 2.41E$-$05 & 2.52E$-$05 & 2.35E$-$05 \\
$^{50}$Ti & 3.12E$-$07 & 2.70E$-$07 & 2.55E$-$07 \\
$^{50}$V  & 3.85E$-$08 & 1.39E$-$08 & 9.37E$-$09 \\
$^{51}$V  & 9.28E$-$05 & 9.53E$-$05 & 8.73E$-$05 \\
$^{50}$Cr & 2.96E$-$04 & 3.02E$-$04 & 2.65E$-$04 \\
$^{52}$Cr & 9.66E$-$03 & 9.87E$-$03 & 9.58E$-$03 \\
$^{53}$Cr & 1.18E$-$03 & 1.18E$-$03 & 1.11E$-$03 \\
$^{54}$Cr & 1.08E$-$05 & 1.06E$-$05 & 1.05E$-$05 \\
$^{55}$Mn & 1.29E$-$02 & 1.27E$-$02 & 1.19E$-$02 \\
$^{54}$Fe & 9.90E$-$02 & 9.54E$-$02 & 8.52E$-$02 \\
$^{56}$Fe & 6.20E$-$01 & 7.19E$-$01 & 8.17E$-$01 \\
$^{57}$Fe & 1.62E$-$02 & 1.86E$-$02 & 2.11E$-$02 \\
$^{58}$Fe & 7.79E$-$05 & 7.55E$-$05 & 7.51E$-$05 \\
$^{59}$Co & 5.09E$-$04 & 5.66E$-$04 & 6.37E$-$04 \\
$^{58}$Ni & 6.27E$-$02 & 6.50E$-$02 & 6.64E$-$02 \\
$^{60}$Ni & 4.49E$-$03 & 5.11E$-$03 & 5.92E$-$03 \\
$^{61}$Ni & 5.33E$-$05 & 6.47E$-$05 & 9.08E$-$05 \\
$^{62}$Ni & 3.05E$-$04 & 3.99E$-$04 & 5.47E$-$04 \\
$^{64}$Ni & 2.53E$-$06 & 8.73E$-$07 & 5.89E$-$07 \\
$^{63}$Cu & 1.17E$-$05 & 4.27E$-$06 & 3.04E$-$06 \\
$^{65}$Cu & 4.28E$-$06 & 1.92E$-$06 & 1.41E$-$06 \\
$^{64}$Zn & 9.90E$-$06 & 7.06E$-$06 & 6.99E$-$06 \\
$^{66}$Zn & 1.34E$-$05 & 8.50E$-$06 & 7.61E$-$06 \\
$^{67}$Zn & 6.23E$-$07 & 2.03E$-$07 & 1.39E$-$07 \\
$^{68}$Zn & 7.56E$-$07 & 3.89E$-$07 & 2.79E$-$07 \\
$^{70}$Zn & 2.06E$-$08 & 6.41E$-$09 & 4.23E$-$09 \\
$^{69}$Ga & 4.16E$-$07 & 1.98E$-$07 & 1.42E$-$07 \\
$^{71}$Ga & 6.61E$-$08 & 2.77E$-$08 & 1.94E$-$08 \\